\documentclass[a4paper,12pt]{article}
\usepackage[english]{babel}
\usepackage[utf8]{inputenc}
\usepackage{t1enc}
\usepackage{floatflt}
\usepackage{graphicx}
\usepackage{psfrag}
\usepackage{bbm}
\usepackage{amsmath}
\usepackage{amssymb}
\usepackage{hyperref}
\usepackage{ifthen}
\usepackage{subfigure}
\usepackage{epstopdf}

\def\d{\mathrm{d}}
\def\e{\mathrm{e}}
\def\imagi{\mathrm{i}}

\def\lag{{\mathcal{L}}}

\def\kihagy#1{}

\newcommand{\arxiv}[2][]{%
  \ifthenelse{\equal{#1}{}}{%
    \href{http://arxiv.org/abs/#2}{\texttt{arXiv:#2}}%
  }{%
    \href{http://arxiv.org/abs/#2}{\texttt{arXiv:#2 [#1]}}%
  }%
}%

\title{Electroweak strings with dark scalar condensates and their (in)stability}
\author{Péter Forgács\textsuperscript{1,2} and Árpád Lukács\textsuperscript{1,3,4}\\
{\small {}\textsuperscript{1} Wigner RCP RMI, H1525 Budapest, POB 49}\\
{\small {}\textsuperscript{2} Institut Denis Poisson CNRS/UMR 7013, Universit\'e de Tours, Parc de Grandmont, 37200 Tours, France}\\
{\small {}\textsuperscript{3} Department of Theoretical Physics, University of the Basque Country UPV/EHU,}\\
{\small POB 644, E-48080 Bilbao, Spain}\\
{\small {}\textsuperscript{4} Kavli Institute of Nanoscience, Lorentzweg 1, 2628CJ Delft, The Netherlands}
}

\begin{document}

\maketitle

\begin{abstract}
The stability of ``visible'' electroweak-type cosmic strings is investigated in an extension of the Standard Model by a minimal dark sector, consisting of a U(1) gauge field, broken spontaneously by a scalar. The ``visible'' and dark sectors are coupled through a Higgs-portal and a gauge-kinetic mixing term. It is found that strings whose core is ``filled'' with a dark scalar condensate exhibit better stability properties than their analogues in the Standard Model, when the electroweak mixing angle is close to $\theta_{\scriptscriptstyle\rm W}=\pi/2$. They become unstable as  one lets $\theta_{\scriptscriptstyle\rm W}$ approach its physical value. The instability mechanism appears to be a W boson condensation mechanism
found in previous studies on the stability of electroweak strings.
\end{abstract}

Cosmic strings are expected to form due to spontaneous symmetry breaking, and have been the subject of vigorous research ever since their first proposition \cite{kibbleorig, VS, kibble, VachaspatiSP, Ringeval}. Since cosmic strings are relics of the phase transitions in the early universe, they may be viewed as a link between high energy physics and cosmology. They are expected to contribute to the anisotropy of the cosmic microwave background \cite{VS, VachaspatiSP, Ringeval, HLUDK} and structure formation \cite{VS, SVL, Duplessis}. At a lower energy scale, % while
electroweak strings may also manifest themselves observationally by creating a primordial magnetic field and play a role in baryogenesis \cite{semilocal}. Cosmic strings exist generically in spontaneously broken gauge theories, the prototype being the Abrikosov-Nielsen-Olesen (ANO) string in the Abelian Higgs model \cite{Abrikosov, NO}. ANO strings can be embedded in the Glashow-Salam-Weinberg (GSW) theory [GSW theory, with its parameters assuming their physical values is the electroweak sector of the Standard Model (SM)] \cite{VachaspatiE, VachaspatiE2, semilocal}.

An important criterion for the relevance of such objects is their stability. In the electroweak theory, embedded cosmic Z-strings are known to have a domain of stability \cite{VachaspatiE, VachaspatiE2, semilocal}. However, it has been found \cite{semilocal, JPV1, JPV2, Perkins, GHelectroweak}, that for physical values of the parameters (more specifically, the electroweak scale, W, Z and Higgs masses), embedded ANO-string solutions are unstable.
The mechanism of the instability is rather transparent in the $\theta_{\scriptscriptstyle\rm W}\to \pi/2$ limit (semilocal model) \cite{vac-ach, hin1, hin2}, in which the Z boson and the Higgs-doublet decouple from the rest of the electroweak theory. The ``extra'' Higgs component condenses into the false vacuum, and thus the string unwinds, the flux is pushed away to infinity.

In theories extending the Standard Model, the possibility arises to ``fill up'' the core of the string, thus preventing the formation of condensates therein. In Ref.\ \cite{FLS}, this possibility has been considered in the case of the semilocal model %(the $\theta_{\scriptscriptstyle\rm W}\to \pi/2$ limit of the GSW theory)
coupled to a dark sector, and a significant enhancement of the stability properties of the string solutions has been found due to the Higgs portal coupling \cite{SilveiraZee, PW} and to gauge kinetic mixing (GKM) \cite{Holdom}. In the present paper, we shall extend this study to the full GSW model coupled to a dark sector.

At this point the following mechanisms for the stabilisation of electroweak strings should be mentioned:
%Other mechanisms for the stabilisation of electroweak strings have been considered recently. We shall mention
additional scalar fields bound in the string \cite{VachaspatiWatkins}, the interplay of quantum fluctuations of neutrinos and deformations of the string \cite{liuvach, grovesperkins, stojk1, stojk2, stojk3}, quantum fluctuations of an additional heavy fermion doublet coupled to the string \cite{Weigel1, Weigel2}, interaction with a thermal photon bath \cite{nagabrand}, and special couplings (of the dilatonic type) \cite{PerivolaropoulosPlatis}.

The model of dark matter we shall consider here is the unified dark matter model put forward in Refs.\ \cite{ArkaniHamed, ArkaniHamed2}, in which it is assumed that in the dark sector there are gauge interactions, the gauge group contains a $U(1)$ factor, which is broken by a dark Higgs field. The dark and the visible sectors interact via the Higgs portal coupling \cite{SilveiraZee, PW} and the GKM \cite{Holdom}. A subset of this model is the scalar phantom dark matter \cite{SilveiraZee, PW}, in which dark matter is scalar, and there is no dark sector gauge field; in this case the dark scalar may have a zero vacuum expectation value. The parameters of the latter model are strongly restricted by observations \cite{Zee1, Zee2, Beniwal}. In the present paper, we consider the case of a non-zero dark scalar vacuum expectation value.
For information on experimental constraints on dark matter, see Ref.\ \cite{PDB}, and in particular, for constraints on the GKM and additional scalar fields, Refs.\ \cite{hook} and \cite{carmi}, respectively.

In the model considered, there exist dark string solutions, i.e., string solutions where the flux is of the dark $U(1)$ interaction, and the dark scalar has a non-zero winding \cite{VachaspatiDS, Vachaspati1, Vachaspati2, Vachaspati3, Holdom2, HindmarshDS, HartmannArbabzadah, BrihayeHartmann, BabeanuHartmann}. Similar solutions in a $U(1)\times U(1)$ theory for higher windings have been considered in Refs.\ \cite{Fidel1, Fidel2}, and an earlier work on string solutions in a portal type theory is Ref.\ \cite{Peter2}. In these works, the strings have a non-zero winding in the dark sector. Dark strings in these models are stable, however, their interactions with the visible sector and their string tension is determined by the (yet unknown) parameters of the dark sector.

The complementary case, in which the flux is in the visible sector, and the role of the dark matter is to stabilise the string, yields a string tension determined by the electroweak scale, and interactions mostly determined by the electroweak parameters.

The semilocal limit of the theory is a generalisation of the Witten model \cite{Witten}, and the string solutions considered in Ref.\ \cite{FLS} are embeddings of the solutions previously found in Refs.\ \cite{Peter, FLCC, FLCC2, Erice}. (Similar and quite interesting string solutions were found in a condensed matter setting, in Refs.\ \cite{BabaevF, BS}.)

In the present paper, we consider the stability of electroweak-dark strings.
We find that the enhanced stability due to the Higgs portal and the GKM couplings found in Ref.\ \cite{FLS} in the full GSW theory coupled to a dark sector only persists to a parameter range of the full theory rather close to the semilocal limit, there extending the domain of stability to $M_{\scriptscriptstyle H}/M_{\scriptscriptstyle Z} > 1$ (up to $M_{\scriptscriptstyle H}/M_{\scriptscriptstyle Z} \sim 1.4$), in contrast to electroweak strings. However, this occurs for parameters disfavoured by experiment, when the dark scalar and the dark Abelian gauge boson are not heavier than their visible counterparts.
We obtain the domain of stability of electroweak-dark strings for various parameter combinations, as well as the dependence of the strength of the instability on the parameters of the dark sector and the strength of the couplings between the visible and the dark sectors.
Our analysis builds upon the results of Refs.\ \cite{Vachaspati1}, \cite{GHelectroweak}.
%We heavily use the analysis of the model (and its dark string solutions) in Ref.\ \cite{Vachaspati1} and that of the stability of electroweak strings in Ref.\ \cite{GHelectroweak}.

The plan of the paper is as follows: we summarise the main characteristics of the model considered in Sec.\ \ref{sec:model}, including the particle content of the model, and the relation among the parameters and the particle properties, based on Ref.\ \cite{Vachaspati1}. Electroweak-dark strings are introduced in Sec.\ \ref{sec:Ansatz}, their stability analysis is performed in Sec.\ \ref{sec:stab}, and we conclude in Sec.\ \ref{sec:discussion}. %The particle content of the model is summarised in Appendices\ \ref{app:phys} and\ \ref{app:scalarmass}, based on Ref.\ \cite{Vachaspati1}.
Some details of the calculations are relegated to Appendix\ \ref{app:details}.

\section{The model considered}\label{sec:model}
We shall consider here string solutions in the GSW model coupled via gauge kinetic mixing \cite{Holdom} and the Higgs portal \cite{SilveiraZee, PW} to a dark sector. The dark sector shall be considered in the unified dark matter model of Refs.\ \cite{ArkaniHamed, ArkaniHamed2, VachaspatiDS}. From the full SM Lagrangian, the terms corresponding to the field that assume non-trivial values in the solutions considered are the electroweak (GSW) and dark sector Abelian gauge terms,
\begin{equation}
 \label{eq:LewG}
 \lag_{G} = -\frac{1}{4}W_{\mu\nu}^a W^{\mu\nu a} -\frac{1}{4}Y_{\mu\nu}Y^{\mu\nu}-\frac{1}{4}C_{\mu\nu}C^{\mu\nu}+\frac{\sin\varepsilon}{2}C_{\mu\nu}Y^{\mu\nu}\,,
\end{equation}
where $W$, $Y$, and $C$ denote the visible sector non-Abelian, Abelian, and the dark sector gauge field strengths, expressed with their respective gauge vector potential as $W_{\mu\nu}^a = \partial_\mu W_\nu^a -\partial_\nu W_\mu^a + g\varepsilon^{abc}W_\mu^b W_\nu^c$, $Y_{\mu\nu} = \partial_\mu Y_\nu-\partial_\mu Y_\nu$, and $C_{\mu\nu} = \partial_\mu C_\nu - \partial_\nu C_\mu$. The fields $W^a_\mu$, $Y_\mu$ and $C_\mu$ are referred to as visible $SU(2)$, $U(1)$ and dark $U(1)$ gauge fields. In the gauge field part of the Lagrangian, Eq.\ (\ref{eq:LewG}), $\varepsilon$ is the gauge kinetic mixing \cite{Holdom, Holdom2}. Its sign is chosen in agreement with Ref.\ \cite{FLS} (and opposite to that of Ref.\ \cite{Vachaspati1}).

Space-time (Greek) indices assume values $\mu,\nu=0,\dots,3$ whereas the internal [SU(2)] indices $a,b,c=1,2,3$. We shall consider the metric $(+,-,-,-)$ and $\varepsilon^{abc}$ is the Levi-Civit\`a symbol.

The scalar sector of the theory consists of the electroweak and the dark Higgs scalars, coupled to their respective gauge fields,
\begin{equation}
 \label{eq:Ls}
 \lag_S = D_\mu \Phi^\dagger D^\mu\Phi + \tilde{D}_\mu \chi^* {\tilde{D}}^\mu\chi -V\,,
\end{equation}
where $D_\mu$ and $\tilde{D}_\mu$ denote the gauge covariant derivatives,
$D_\mu\Phi = \left(\partial_\mu - \frac{\imagi g}{2}W_\mu^a \tau^a -\frac{\imagi g'}{2} Y_\mu\right)\Phi$ and $\tilde{D}_\mu \chi = \left(\partial_\mu -\frac{\imagi \hat{g}}{2}C_\mu\right) \chi$, $\dagger$ denotes adjoint (transposed complex conjugate) and $*$ complex conjugate, and $\tau^a$ are the Pauli matrices in internal (isospin) space. The potential is
\begin{equation}
 \label{eq:pot}
 V=\lambda_1(\Phi^\dagger\Phi-\eta_1^2)^2 + \lambda_2 (|\chi|^2-\eta_2^2)^2+\lambda'(\Phi^\dagger\Phi-\eta_1^2)(|\chi|^2-\eta_2^2)\,.
\end{equation}
%For more details, and the definitions of the field strength tensors and the gauge covariant derivatives, see Appendix \ref{app:defi}.
The Lagrangians (\ref{eq:LewG}) and (\ref{eq:Ls}) %expressed with these fields
reflect the symmetries of the model in a manifest form. On the other hand, the particle content of the theory is better expressed with the so-called physical fields, for which, see Subsec.\ \ref{ssec:ptcle}.

\subsection{Particle content and physical parameters}\label{ssec:ptcle}
Let us briefly consider the particle content and the parameters of the theory, following the analysis in Ref.\ \cite{Vachaspati1}.

To identify physical degrees of freedom, one needs to introduce new fields with the transformation
\begin{equation}\label{eq:gaugediag}
 \begin{pmatrix} Y_\mu \\ W_\mu^3 \\ C_\mu \end{pmatrix}
 = {\bf M}
 \begin{pmatrix} A_\mu \\ Z_\mu \\ X_\mu \end{pmatrix}\,,
\end{equation}
where the (non-unitary) matrix of the transformation is
\begin{equation}
 \label{eq:MMat}
 {\bf M} = \begin{pmatrix}
            c_{\rm W} & -s_{\rm W} c_\zeta        &  s_{\rm W} s_\zeta + t_\varepsilon c_\zeta\\
            s_{\rm W} &  c_{\rm W} c_\zeta        & -c_{\rm W} s_\zeta\\
            0   &  s_\zeta/c_\varepsilon & c_\zeta/c_\varepsilon\\
           \end{pmatrix}\,,
\end{equation}
where $c_{\rm W}=\cos\theta_{\scriptscriptstyle\rm W}$, $s_{\rm w}=\sin\theta_{\scriptscriptstyle\rm W}$, and where $\theta_{\scriptscriptstyle\rm W}$ is the Weinberg angle, $\tan\theta_{\scriptscriptstyle\rm W}=g'/g$,
$c_\varepsilon=\cos\varepsilon$, $s_\varepsilon=\sin\varepsilon$, $t_\varepsilon=s_\varepsilon/c_\varepsilon$, $c_\zeta=\cos\zeta$, and $s_\zeta=\sin\zeta$. The angle $\zeta$ is defined by
\begin{equation}
 \label{eq:zeta}
 \tan2\zeta=\frac{2\sin\theta_{\scriptscriptstyle\rm W}\sin\varepsilon \cos\varepsilon}{R^2-1+\sin^2\varepsilon(1+\sin^2\theta_{\scriptscriptstyle\rm W})}\,,
\end{equation}
and $R=\hat{g}\eta_2/(\bar{g}\eta_1)$ and $\bar{g}=\sqrt{g^2+{g'}^2}$.

In what follows, we shall denote the middle line of the matrix ${\bf M}$ by $\vec{\alpha}$, i.e.,
\begin{equation}
 \label{eq:W3}
 W^3_\mu = \alpha_1 A_\mu + \alpha_2 Z_\mu + \alpha_3 X_\mu\,,\quad \alpha_1=s_w\,,\ \alpha_2=c_W c_\zeta\,,\ \alpha_3=-c_W s_\zeta\,.
\end{equation}
In the new variables, the gauge Lagrangian can be recast as
\begin{equation}
 \label{eq:LewG2}
 \begin{aligned}
 \lag_G = &-\frac{1}{4}F_{\mu\nu}F^{\mu\nu}-\frac{1}{4}Z_{\mu\nu}Z^{\mu\nu}-\frac{1}{4}X_{\mu\nu}X^{\mu\nu}-\frac{1}{4}\tilde{W}_{\mu\nu}{\tilde{W}}^{\mu\nu}\\
 &-g {\tilde{W}}_{\mu\nu}^3 W^{\mu1}W^{\nu2} -g {\tilde{W}}_{\mu\nu}^1 W^{\mu2}W^{\nu3}+g {\tilde{W}}_{\mu\nu}^2 W^{\mu1}W^{\nu3}\\
 &-\frac{g^2}{4}(W_{\mu}^a W^{\mu a})^2 -\frac{g^2}{2}W_\mu^3 W^{\mu 3}W_\nu^a W^{\nu a} +\frac{g^2}{4}W_\mu^a W_\nu^a W^{\mu b}W^{\nu b}-\frac{g^2}{2}W_\mu^3 W_\nu^3 W^{\mu a}W^{\nu a}\,,
\end{aligned}
\end{equation}
where in Eq.\ (\ref{eq:LewG2}) $a=1,2$ and ${\tilde{W}}_{\mu\nu}^a=\partial_\mu W_\nu^a - \partial_\nu W_\mu^a$ (i.e., the linear part of the field strength tensor).  Eq.\ (\ref{eq:LewG2}) shows that the transformation (\ref{eq:gaugediag}) results in decoupled kinetic and mass terms for the new vector fields $A_\mu$, $Z_\mu$, and $X_\mu$.

In the scalar sector, the particles correspond to amplitude fluctuations of the Higgs field assuming a vacuum expectation value ($\phi_2$) and, similarly, amplitude fluctuations of the dark scalar $\chi$ \cite{Vachaspati1}. Here we convert the formulae of Ref.\ \cite{Vachaspati1} to our notations for convenience. The scalar mass matrix in the basis of the fields $h=\sqrt{2}(|\phi_1|-\eta_1)$ and $s=\sqrt{2}(|\chi|-\eta_2)$ in the Lagrangian (\ref{eq:Ls}) is
\[
 \begin{pmatrix}
  m_{\scriptscriptstyle H}^2 & 2\lambda' \eta_1 \eta_2\\
  2\lambda' \eta_1 \eta_2 & m_{\scriptscriptstyle S}^2
 \end{pmatrix}\,,
\]
where $m_{\scriptscriptstyle H}^2=4\lambda_1 \eta_1^2$ and $m_{\scriptscriptstyle S}^2=4\lambda_2 \eta_2^2$. The physical fields $\phi_{\scriptscriptstyle H}$, $\phi_{\scriptscriptstyle S}$ are rotated at an angle $\phi_s$,
\begin{equation}\label{eq:scalarfields}
 \begin{pmatrix} h\\s \end{pmatrix}
=\begin{pmatrix}
   \cos\theta_s & \sin\theta_s\\
  -\sin\theta_s & \cos\theta_s
 \end{pmatrix}
 \begin{pmatrix} \phi_{\scriptscriptstyle H}\\ \phi_{\scriptscriptstyle S} \end{pmatrix}
\end{equation}
and the scalar mixing angle is given as
\begin{equation}
 \label{eq:scalarangle}
 \tan2\theta_s = \frac{4\lambda' \eta_1\eta_2}{4\lambda_2 \eta_2^2-4\lambda_1\eta_1^2}
 %= \frac{4\beta' \eta_{2s}}{2\beta_2 \eta_{2s}^2-2\beta_1}\,.
\end{equation}
The corresponding eigenvalues (squared scalar masses) are
\begin{equation}
 \label{eq:scalarmasses}
 \begin{aligned}
 M_{\scriptscriptstyle H}^2&=m_{\scriptscriptstyle H}^2-(m_{\scriptscriptstyle S}^2-m_{\scriptscriptstyle H}^2)\frac{\sin^2\theta_s}{\cos2\theta_s}\,,\\
 M_{\scriptscriptstyle S}^2&=m_{\scriptscriptstyle S}^2+(m_{\scriptscriptstyle S}^2-m_{\scriptscriptstyle H}^2)\frac{\sin^2\theta_s}{\cos2\theta_s}\,.\\
 \end{aligned}
\end{equation}
For more details, see Ref.\ \cite{Vachaspati1}.

The couplings of the physical fields  are calculated in Ref.\ \cite{Vachaspati1}; which are reproduced here with the replacement $\varepsilon\to-\varepsilon$ (for agreement with Ref.\ \cite{FLS}):
\begin{equation}\label{eq:couplings}
\begin{aligned}
 g_{\scriptscriptstyle A\phi^+} &= e\,,\\
 g_{\scriptscriptstyle Z\phi^+} &= c_\zeta \frac{e}{2}\left(\frac{1}{t_{\rm W}}-t_{\rm W}\right)+s_\zeta \frac{e}{2}\frac{t_\varepsilon}{c_{\rm W}}\,,\\
 g_{\scriptscriptstyle X\phi^+} &= c_\zeta \frac{e}{2}\frac{t_\varepsilon}{c_{\rm W}}-s_\zeta\frac{e}{2}\left(\frac{1}{t_{\rm W}}-t_{\rm W}\right)\,,\\
 g_{\scriptscriptstyle AH} &=0\,,\\
 g_{\scriptscriptstyle ZH} &= -c_\zeta\frac{e}{2}\frac{1}{s_{\rm W}c_{\rm W}}+s_\zeta \frac{e}{2}\frac{t_\varepsilon}{c_{\rm W}}\,,\\
 g_{\scriptscriptstyle XH} &= c_\zeta \frac{e}{2}\frac{t_\varepsilon}{c_{\rm W}}+s_\zeta \frac{e}{2}\frac{1}{s_{\rm W}c_{\rm W}}\,,\\
 g_{\scriptscriptstyle AS} &= 0\,,\\
 g_{\scriptscriptstyle ZS} &= s_\zeta \frac{\hat{g}}{2}\frac{1}{c_\varepsilon}\,,\\
 g_{\scriptscriptstyle XS} &= c_\zeta \frac{\hat{g}}{2}\frac{1}{c_\varepsilon}\,.
\end{aligned}
\end{equation}

The gauge covariant derivatives of the scalars expressed with the physical gauge fields and the couplings from Eq. (\ref{eq:couplings}) are
\begin{equation}
 \label{eq:covariant2}
 D_\mu \Phi = \begin{pmatrix}
               (\partial_\mu - \imagi g_{\scriptscriptstyle A\phi^+}A_\mu - \imagi g_{\scriptscriptstyle Z\phi^+}Z_\mu -\imagi g_{\scriptscriptstyle X\phi^+}X_\mu)\phi_1 -\frac{\imagi g}{\sqrt{2}}W_\mu^+ \phi_2\\
               (\partial_\mu - \imagi g_{\scriptscriptstyle AH}A_\mu - \imagi g_{\scriptscriptstyle ZH}Z_\mu -\imagi g_{\scriptscriptstyle XH}X_\mu)\phi_2 -\frac{\imagi g}{\sqrt{2}}W_\mu^- \phi_1
              \end{pmatrix}\,,
\end{equation}
where $W_\mu^{\pm}=\frac{1}{\sqrt{2}}(W_\mu^1 \mp \imagi W_\mu^2)$, $g_{\scriptscriptstyle AH}=0$, and
\begin{equation}
 \label{eq:covariant3}
 \tilde{D}_\mu \chi = (\partial_\mu -\imagi g_{\scriptscriptstyle AS}A_\mu -\imagi g_{\scriptscriptstyle ZS}Z_\mu -\imagi g_{\scriptscriptstyle XS}X_\mu)\chi\,.
\end{equation}
Note, that $g_{\scriptscriptstyle AS}=0$, i.e., the dark scalar is indeed dark.

The vector boson masses are
\begin{equation}
 \label{eq:gaugemasses}
 M_{\scriptscriptstyle W}^2 = \frac{g^2\eta_1^2}{2}\,,\quad M_{\scriptscriptstyle Z}^2 =  2 g_{\scriptscriptstyle ZH}^2 \eta_1^2 + 2 g_{\scriptscriptstyle ZS}^2\eta_2^2\,,\quad M_{\scriptscriptstyle X}^2 = 2 g_{\scriptscriptstyle XH}^2 \eta_1^2 + 2 g_{\scriptscriptstyle XS}^2\eta_2^2\,.
\end{equation}
For more details, see Ref.\ \cite{Vachaspati1}.

The $g \to 0$ ($\theta_{\scriptscriptstyle\rm W} \to \pi/2$) limit is referred to as the semilocal limit; in particular that limit of the model is the semilocal-dark model. In this limit, the non-Abelian gauge field decouples, and the $SU(2)$ symmetry becomes global.

\subsection{Values of model parameters considered}\label{ssec: exp}
The parameters of the visible sector, the electroweak parameters have been determined to a high accuracy \cite{PDB}. In what follows, for a solution to be considered physical, setting electroweak parameters (W and Z masses, electric charge, and Weinberg angle) to their physical value is considered necessary.

The dark sector gauge $M_{\scriptscriptstyle X}$ and scalar $M_{\scriptscriptstyle S}$ masses are experimentally bound to be larger than their visible sector counterparts, to avoid abundant dark decays, unless the coupling between the visible and dark sectors is extremely weak. The scalar mixing angle $\theta_s$ is largely unconstrained as long as the dark sector particles are heavy enough \cite{ArkaniHamed, ArkaniHamed2}.

For observational bounds on the model parameters, see Ref.\ \cite{hook} for those on the GKM, Ref.\ \cite{carmi} for those on the scalar sector, and Ref.\ \cite{PDB} for a review. For our purposes, it shall be sufficient to know, that for $M_{\scriptscriptstyle X} < 200\,{\rm GeV}$, $|\varepsilon| \lesssim 0.03$ (and for a large part of the dark gauge boson mass range, $|\varepsilon| \lesssim 10^{-3}$),
and that $|\theta_s| < \pi/2$.
%, and that the squared sine of the scalar mixing angle is also $\lesssim 0.5$.
For heavy dark sector particles, the model is largely unconstrained \cite{Vachaspati1}.

\subsection{Rescaling}\label{ssec:rescale}
For simplicity sake, we shall also rescale the coordinates and the fields as $\Phi \to \eta_1 \Phi$, $\chi\to \eta_1 \chi$ and $x^\mu \to x^\mu / (g_{\scriptscriptstyle ZH}\eta_1)$. All gauge couplings will be rescaled by a factor of $g_{\scriptscriptstyle ZH}$, i.e., one shall perform the replacement $\eta_1\to 1$, $\eta_2 \to \eta_{2s}=\eta_2/\eta_1$, $g_{\scriptscriptstyle ZH}\to 1$, $g_{\scriptscriptstyle XH}\to g_{\scriptscriptstyle XHs} = g_{\scriptscriptstyle XH}/g_{\scriptscriptstyle ZH}$, etc. We shall
introduce the notation $\beta_{1,2} = 2\lambda_{1,2}/g_{\scriptscriptstyle ZH}^2$ and $\beta'=\lambda'/g_{\scriptscriptstyle ZH}^2$, the analogues of the Ginzburg-Landau parameter $\beta=M_{\scriptscriptstyle H}/M_{\scriptscriptstyle Z}$ in the GSW model. When no confusion is possible, the subscript ``s'' shall be dropped.
%The scalar mixing angle is expressed with the rescaled parameters as $\tan2\theta_s= 4\beta' \eta_{2s}/({2\beta_2 \eta_{2s}^2-2\beta_1})$.

%For details of the physical fields in the gauge sector, and their couplings to the scalars, see Ref.\ \cite{Vachaspati1}, and Appendix \ref{app:phys}. Also in Appendix \ref{app:phys}, we consider
The rescaled parameters $\beta_{1,2}$ and $\beta'$ (coefficients of the quartic terms in the rescaled potential) play somewhat analogous roles in the radial equations of cylindrically symmetric strings as the ratio of the scalar and the vector masses in the Abelian Higgs and semilocal models, and the ratio of the Higgs and Z boson masses in the GSW model, $\beta=M_{\scriptscriptstyle H}/M_{\scriptscriptstyle Z}$, which we shall refer to as the Ginzburg-Landau parameter. For the coupled electroweak-dark sector, no such simple relation between the mass ratio and the rescaled potential parameters is known.
%Rescalings, and analogues of the Ginzburg-Landau parameter, $\beta=M_{\scriptscriptstyle H}/M_{\scriptscriptstyle Z}$ of the SM are also given there.
%For the physical fields in the scalar sector, see Appendix \ref{app:scalarmass}.

\section{Electroweak-dark strings}\label{sec:Ansatz}
The ANO-string \cite{Abrikosov,NO} is a well-known cylindrically symmetric solution of the Abelian Higgs model, in which the scalar field has a winding number $n$, the gauge field has a non-vanishing radial component, and the resulting string or flux tube contains $n$ flux quanta.

The ANO string can be embedded in the GSW theory by assuming that the component of the Higgs field having non-zero expectation value in the vacuum has a winding, and the flux is in the Z field. Using cylindrical coordinates $r,\vartheta,z$, the Ansatz
\begin{equation}\label{eq:Ansatz1}
  \phi_2  = f(r)\e^{\imagi n \vartheta}\,,\quad\quad
Z_\vartheta = n {\mathfrak{z}}(r)\,,
\end{equation}
describes a cylindrically symmetric vortex string (or flux tube)  centred on the $z$-axis, with $n$ flux quanta\cite{VS,kibble,semilocal}.

The unified dark matter model \cite{ArkaniHamed, ArkaniHamed2} extends the GSW model with a dark sector, containing a Higgs field $\chi$ and an additional $U(1)$ gauge field. The Ansatz (\ref{eq:Ansatz1}) is accordingly extended, preserving cylindrical symmetry, as
\begin{equation}
 \label{eq:Ansatz2}
 \chi = f_d(r)\,,\quad\quad X_\vartheta = n x(r)\,,
\end{equation}
where the fields $Z$ and $X$ are the physical fields obtained from a combination of $Y$ and $X$.

\subsection{Radial equations of the vortex solutions}\label{app:bgrradeq}
Plugging in the Ansatz (\ref{eq:Ansatz1}), (\ref{eq:Ansatz2}) into the field equations yields the radial equations,
\begin{equation}
 \label{eq:radeq}
 \begin{aligned}
  \frac{1}{r}(rf')' &= f\left[\frac{n^2(1-\mathfrak{z}-g_{\scriptscriptstyle XH}x)^2}{r^2}+\beta_1(f^2-1)+\beta'(f_d^2-\eta_2^2)\right]\,,\\
  \frac{1}{r}(rf_d')' &= f_d\left[\frac{n^2(g_{\scriptscriptstyle ZS}\mathfrak{z}-g_{\scriptscriptstyle XS}x)^2}{r^2}+\beta_2(f_d^2-\eta_2^2)+\beta'(f^2-1)\right]\,,\\
  r(\mathfrak{z}'/r)' &= 2 f^2(\mathfrak{z}+g_{\scriptscriptstyle XH}x-1) + 2 g_{\scriptscriptstyle ZS}f_d^2(g_{\scriptscriptstyle ZS}\mathfrak{z}+g_{\scriptscriptstyle XS}x)\,,\\
  r(x'/r)' &= 2 g_{\scriptscriptstyle XH}f^2(\mathfrak{z}+g_{\scriptscriptstyle XH}x-1) + 2 g_{\scriptscriptstyle XS}f_d^2(g_{\scriptscriptstyle ZS}\mathfrak{z}+g_{\scriptscriptstyle XS}x)\,,
 \end{aligned}
\end{equation}
where a prime on the radial functions (but not on the constant $\beta'$) denotes $\d/\d r$, and $r$ denotes the (rescaled) radial coordinate. Note, that without the dark sector, one would get the ANO vortex \cite{Abrikosov, NO} embedded in the $Z$ field.

The energy density of a field configuration within the Ansatz (\ref{eq:Ansatz1}), (\ref{eq:Ansatz2}) is
\begin{equation}
 \label{eq:Edens}
 \begin{aligned}
 \mathcal{E} = &\frac{n^2}{2}\left[ \left(\frac{\mathfrak{z}'}{r}\right)^2+\left(\frac{x'}{r}\right)^2\right]+(f')^2+(f_d')^2\\
 &+\frac{n^2(1-\mathfrak{z}-g_{\scriptscriptstyle XH}x)^2f^2}{r^2}+\frac{n^2(g_{\scriptscriptstyle ZS}\mathfrak{z}-g_{\scriptscriptstyle XS}x)^2f_d^2}{r^2}+V\,,
\end{aligned}
\end{equation}
where $V=\beta_1(f^2-1)^2/2+\beta_2 (f_d^2-\eta_2^2)^2/2+\beta'(f^2-1)(f_d^2-\eta_2^2)$ with $\beta_i = 2\lambda_i/g_{\scriptscriptstyle ZH}^2$ and $\beta'=\lambda'/g_{\scriptscriptstyle ZH}^2$, is the (rescaled) potential. The energy within a given radius is $E(r)=2\pi \int_0^r \mathcal{E}r\d r$.

\subsection{Electroweak, semilocal, and dark strings}
In the Abelian Higgs model, ANO strings are topologically stable. Note, that for embeddings of ANO strings to an enlarged model, new instabilities may arise which excite the additional fields, therefore embedded ANO vortices, may become unstable.
Semilocal strings given with Ansatz (\ref{eq:Ansatz1}) in the semilocal model correspond to embedded ANO strings. Their stability depends now on the Ginzburg-Landau parameter. For $\beta < 1$, the simplest $n=1$ semilocal strings are stable, and become unstable for $\beta >1$ \cite{hin1, hin2, semilocal}. The mechanism of the instability is that a condensate of the other Higgs component, $\phi_1$ forms in the core of the string, and eventually dilutes the flux.

%The $\theta_{\scriptscriptstyle\rm W}\to \pi/2$ limit of the GSW model is referred to as the semilocal model; it is the Abelian Higgs model extended with a second Higgs component (the non-Abelian gauge field decouples, and the $SU(2)$ symmetry becomes global).
%String solutions in the semilocal model within the Ansatz (\ref{eq:Ansatz1}) are termed semilocal strings; they are embedded ANO strings, and their stability depends on the Ginzburg-Landau parameter also for $n=1$. For $\beta < 1$, semilocal strings are stable, and unstable for $\beta >1$ \cite{hin1, hin2, semilocal}. The mechanism of the instability is that a condensate of the other Higgs component, $\phi_1$ forms in the core of the string, and eventually dilutes the flux.

In the GSW model, strings within the Ansatz (\ref{eq:Ansatz1}) are referred to as electroweak strings or Z-strings. Their stability depends on the parameters of the model. They are stable for $\beta=M_{\scriptscriptstyle H}/M_{\scriptscriptstyle Z} < 1$ and  for values of the Weinberg angle $\theta_{\scriptscriptstyle\rm W}$ close to $\pi/2$; i.e., they are only stable close to the semilocal limit \cite{JPV1, JPV2, Perkins, GHelectroweak}. The mechanism of the instability is unwinding through the condensation of Higgs and W bosons in the string core.

In the model outlined above and its semilocal ($\theta_{\scriptscriptstyle\rm W}\to \pi/2$)
limit, string solutions with winding in the dark sector have been considered in Refs.\ \cite{VachaspatiDS, HartmannArbabzadah, BrihayeHartmann, BabeanuHartmann}. These strings are topologically stable. Their energy scale is determined by the scale of the symmetry breaking in the dark sector, which is presently to a large extent unconstrained by measurement.

The scale of strings in the visible sector, within the Ansatz (\ref{eq:Ansatz1}) and (\ref{eq:Ansatz2}), is the electroweak scale. This is the main motivation behind the search for mechanisms stabilising electroweak strings. Besides, as the mechanism behind the instability is the formation of condensates in the string core, the idea arises naturally to look for other fields which may fill up the core, thus preventing the instability. In Ref.\ \cite{FLS}, this idea has been considered in the semilocal limit of the model considered here, i.e., in the semilocal model extended with a scalar and another $U(1)$ gauge field in the dark sector. There, two cases
%, termed 1VEV and 2VEV
have been considered, depending on whether only the visible Higgs or both the visible and the dark scalar field obtain a vacuum expectation value. Relevant to the dark matter model of Refs.\ \cite{ArkaniHamed, ArkaniHamed2} is the latter case. In both cases it has been found in Ref.\ \cite{FLS} that the stabilising effect is significant, semilocal-dark strings may exist for $\beta$ significantly above unity.

The semilocal-dark strings of Ref.\ \cite{FLS} in the case with no GKM may be considered embeddings of string solutions in non-symmetric extended Abelian Higgs models considered in Refs.\ \cite{FLCC, FLCC2}. Also, in the %1VEV 
case,
where only the visible sector scalar obtains a vacuum expectation value,
semilocal and semilocal-dark strings may coexist, and their stability is considered separately. The energy of semilocal-dark strings is lower, and they are stable for a larger set of parameters.

\begin{figure}[h!]
 \noindent\hfil\includegraphics{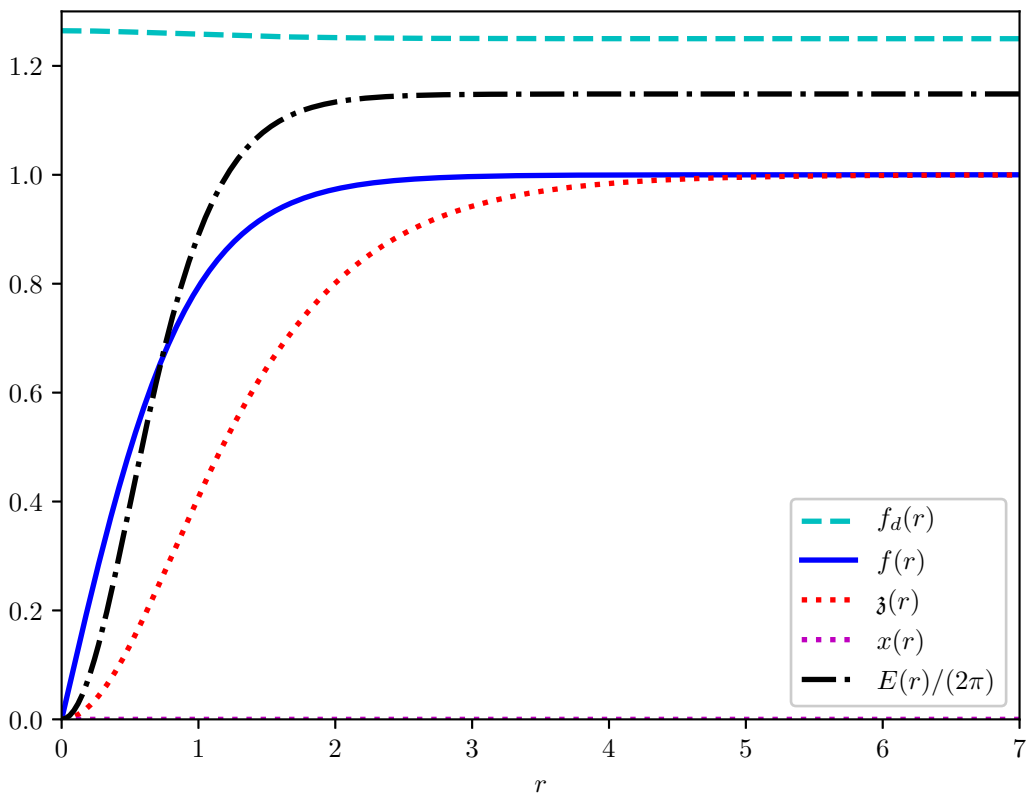}
 \caption{Radial profile functions of an electroweak-dark string. The visible sector parameters are set to their physical values, $M_{\scriptscriptstyle Z}=80.4\,{\rm GeV}$, $M_{\scriptscriptstyle W}=91.2\,{\rm GeV}$, $M_{\scriptscriptstyle H}=125.1\,{\rm GeV}$, $e=0.3086$, the dark sector parameters are $M_{\scriptscriptstyle S}=132.8\,{\rm GeV}$, $g_{\scriptscriptstyle XS}=0.3086$, and the scalar mixing angle is $\theta_s=0.51$. The dark sector charge of the Higgs is $g_{\scriptscriptstyle X\phi^+}=0$ (no GKM, $\varepsilon=0$). For these parameter values, $x(r)=0$.
 % An example vortex background, physical SM parameters, $M_{\scriptscriptstyle X}=94.8\,{\rm GeV}$, $M_{\scriptscriptstyle S}=141.42\,{\rm GeV}$, $\theta_s=0.501$, $g_{\scriptscriptstyle X \phi^+}/g_{\scriptscriptstyle ZH}=2\cdot 10^{-3}$.
 }
 \label{fig:bgrxample}
\end{figure}

Here, we consider electroweak-dark strings, i.e., solutions within the Ansatz (\ref{eq:Ansatz1}), (\ref{eq:Ansatz2}) within the full GSW model coupled to a dark sector containing an $U(1)$ gauge field and a scalar. The resulting radial equations are given in Subsec.\ \ref{app:bgrradeq}. The solutions are found using the shooting to a fitting point method \cite{numrec}, and an example is displayed in Fig.\ \ref{fig:bgrxample}. SM parameters are set to physical values, and dark sector parameters are set to such values, that they are heavier than their visible counterparts. In addition to the profile functions $f$, $f_d$, $\mathfrak{z}$, and $x$, the energy within a radius is shown. (For SM parameter values, see Ref.\ \cite{PDB}.)

\section{Stability analysis}\label{sec:stab}
We analyse the stability of the electroweak-dark strings by linearising the field equations around them.
The linear perturbations added to the fields are denoted by
\begin{equation}\label{eq:pertPsi}
\Psi=(\delta A_\mu, \delta W_\mu^\pm, \delta Z_\mu,\delta X_\mu,\delta \phi_a, \delta \chi)\,.
\end{equation}
Since the string is electrically neutral, the electromagnetic field perturbations $\delta A_\mu$ decouple and satisfy a free wave equation, i.e., they play no role in the stability of the string. The string possesses a global direction in internal space [$\phi_1=0$ in Eqs.\ (\ref{eq:Ansatz1}), (\ref{eq:Ansatz2})], which results in further decouplings. It turns out that there are four decoupled blocks:
\begin{equation}\label{eq:blocks}
\begin{aligned}
({\rm i})\ &\Psi^{\rm i}=(\delta A_\mu)\,,\\\
({\rm ii})\ &\Psi^{\rm ii}=(\delta W^+_\mu, \delta\phi_1)\,,\\
\end{aligned}\quad\quad
\begin{aligned}
({\rm iii})\ &\Psi^{\rm iii}=(\delta W^-_\mu, \delta \phi_1^*)\,,\\
({\rm iv})\ &\Psi^{\rm iv}=(\delta Z_\mu, \delta X_\mu, \delta\phi_2, \delta\phi_2^*, \delta\chi, \delta\chi^*)\,,\\
\end{aligned}
\end{equation}
each block satisfying an equation of the form
\begin{equation}\label{eq:lineqM}
 \mathcal{D}^I \Psi^I = 0\,,\ I={\rm i},\ldots{\rm iv}\,,
\end{equation}
where $\mathcal{D}^I$ is a matrix with differential operators in the diagonal and coupling terms in the remaining elements. Of the four blocks, (iii) is the conjugate of (ii) and, therefore, admits the same (real) eigenvalues.

%The gauge condition used for the background may also receive first order corrections.
To ensure that the linearised equations (\ref{eq:lineqM}) have properties suitable for our numerical solution procedure, we find that an appropriate gauge choice for the perturbations is the background field gauge \cite{Goodband, GHelectroweak}, which is defined as
\begin{equation}
 \label{eq:gaugechoiceM}
 \begin{aligned}
  F_1 &= \partial_\mu \delta W^{\mu+} -\imagi g W_\mu^3 \delta W^{\mu+}-\frac{\imagi g}{\sqrt{2}}\phi_2^*\delta\phi_1=0\,,\\
  F_2 &= \partial_\mu \delta W^{\mu-} +\imagi g W_\mu^3 \delta W^{\mu-}+\frac{\imagi g}{\sqrt{2}}\phi_2\delta\phi_1^*=0\,,\\
  F_3 &= \partial_\mu\delta Z^\mu + \imagi g_{\scriptscriptstyle ZH}(\phi_2 \delta\phi_2^*-\phi_2^* \delta\phi_2)
  +\imagi g_{\scriptscriptstyle ZS}(\chi \delta\chi^*-\chi^* \delta\chi)=0\,,\\
  F_4 &= \partial_\mu\delta X^\mu + \imagi g_{\scriptscriptstyle XH}(\phi_2 \delta\phi_2^*-\phi_2^* \delta\phi_2)
  +\imagi g_{\scriptscriptstyle XS}(\chi \delta\chi^*-\chi^* \delta\chi)=0\,.
 \end{aligned}
\end{equation}
Gauge conditions (\ref{eq:gaugechoiceM}) are imposed by adding the gauge fixing terms $\sum_i|F_i|^2/2$ to the second order terms of the Lagrangian. In the fluctuation equations, they cancel the first order derivative terms, and the time derivatives are readily isolated \cite{Goodband, GHelectroweak}. (Note, that some gauge degrees of freedom still remain, satisfying ``ghost''-equations, which all have positive eigenvalues.)

We shall now follow the treatment of Refs.\ \cite{Goodband, GHelectroweak, FL, FLCC, FLCC2, FLS} to bring the perturbation equations to a form suitable for numerical solutions. For more details, as well as for the full set of linearised equations, we refer to Appendix\ \ref{app:details}.

Because of the time- and $z$-coordinate independence of the string solution, the corresponding fluctuation equations of the gauge fields decouple further. The equations of the temporal and the $z$ components of the gauge fields do not contribute to the instabilities (see Appendix \ref{app:details} for details).

This $t,z$-independence of the background solution can be further exploited by separating harmonic components of the perturbations, i.e., assuming a time-dependence of the form $\Psi^I = \exp[i(\Omega t-kz)]\Phi^I$, transforming Eq.\ (\ref{eq:lineqM}) into
\begin{equation}\label{eq:lineqF}
 D^I \Phi^I = (\Omega^2-k^2)\Phi^I\,,
\end{equation}
where an eigenvalue $\Omega^2 < 0$ signals instability, and $D^I$ is a matrix of differential operators (the spatial part of $\mathcal{D}^I$). The lowest eigenvalue corresponds to $k=0$, therefore, in what follows, this $k=0$ is considered.

Because of the cylindrical symmetry of the string, Eq.\ (\ref{eq:lineqF}) can be reduced to ordinary differential equations by the Fourier transformation in the angular coordinate $\vartheta$, reducing Eq.\ (\ref{eq:lineqF}) to
\begin{equation}\label{eq:radeqEigM}
 \mathcal{M}^{I}_\ell \Phi^I_\ell = \Omega^2 \Phi^I_\ell\,.
\end{equation}

The known instabilities of the electroweak strings are in the sector of the perturbations consisting of the fields $W^+$ and $\delta\phi_1$ (or equivalently $W^-$ and $\delta\phi_1^*$) \cite{GHelectroweak, VS, kibble}. The remaining sectors are deformations of their counterparts in the case of ANO strings, and thus not expected to contain further instabilities (as the corresponding blocks for the ANO string have large positive eigenvalues).

In sector (ii) $\Psi^{\rm ii}=(\delta W^+_\mu, \delta\phi_1)$. The Fourier transformation singles out a mode of the form
\begin{equation}\label{eq:pawaveR}
\begin{aligned}
 \delta \phi_1 &= s_{1,\ell}(r) \e^{\imagi \ell \vartheta}\e^{\imagi\Omega t}\,,\\
 \delta W^+_+ &= \imagi w_{+,\ell}(r) \e^{\imagi (\ell-1-n)\vartheta}\e^{\imagi\Omega t}\,,\\
 \delta W^+_- &= -\imagi w_{-,\ell}(r) \e^{\imagi (\ell+1-n)\vartheta}\e^{\imagi\Omega t}\,,\\
\end{aligned}
\end{equation}
where $\delta W^\pm_{+} = \exp(-\imagi\vartheta)(\delta W^\pm_r -\imagi \delta W^\pm_\vartheta/r)$, $\delta W^\pm_-=\delta W^\pm_+{}^*$, i.e., $\Psi_\ell = (s_{1,\ell}$, $w_{+,\ell}$, $w_{-,\ell}^*)$.
The matrix operator of Eq.\ (\ref{eq:radeqEigM}) in this block is
\begin{equation}
 \label{eq:radeqPMM}
 \mathcal{M}_{\ell}=\begin{pmatrix}
                              D_{\ell,1}  & B_{1+,\ell} & B_{1-,\ell}\\
                              B_{1+,\ell} & D_{+,\ell} & 0          \\
                              B_{1-,\ell} & 0          & D_{-,\ell}
                             \end{pmatrix}\,,
\end{equation}
where
\begin{equation}
 \label{eq:radeqPOM}
\begin{aligned}
 D_{\ell,1} &= -\frac{1}{r}\frac{\d}{\d r}r\frac{\d}{\d r} + \left( \frac{[n(g_{\scriptscriptstyle Z\phi^+}\mathfrak{z}+g_{\scriptscriptstyle X\phi^+}x)-\ell]^2}{r^2}+\beta_1 (f^2-1)+\beta'(f_d^2-\eta_2^2)+\frac{g^2 }{2}f^2\right)\,,\\
 % \eta_1^2
 D_{+,\ell} &= -\frac{1}{r}\frac{\d}{\d r}r\frac{\d}{\d r} + \left( \frac{[\ell-1-n(1+g(\alpha_2 \mathfrak{z} +\alpha_3 x))]^2}{r^2}+\frac{g^2}{2}f^2-2\frac{gn}{r}(\alpha_2 \mathfrak{z}'+\alpha_3 x')\right)\,,\\
 D_{-,\ell} &= -\frac{1}{r}\frac{\d}{\d r}r\frac{\d}{\d r} + \left( \frac{[\ell+1-n(1+g(\alpha_2 \mathfrak{z} +\alpha_3 x))]^2}{r^2}+\frac{g^2}{2}f^2+2\frac{gn}{r}(\alpha_2 \mathfrak{z}'+\alpha_3 x')\right)\,,\\
\end{aligned}
\end{equation}
and
\begin{equation}
 \label{eq:radeqPCM}
 \begin{aligned}
 B_{1+,\ell} &= -g \left( f'-\frac{n f}{r}(1-g_{\scriptscriptstyle ZH}  \mathfrak{z}-g_{\scriptscriptstyle XH}x)\right)\,,\\
 B_{1-,\ell} &= \ \ g \left( f'+\frac{n f}{r}(1-g_{\scriptscriptstyle ZH}  \mathfrak{z}-g_{\scriptscriptstyle XH}x)\right)\,.\\
 \end{aligned}
\end{equation}
%An eigenvalue $\Omega^2 < 0$ signals instability (an exponentially growing mode).
The negative eigenvalue for the unit flux $n=1$ string considered here is found in the $\ell=0$ sector.

In Eq.\ (\ref{eq:radeqPMM}) in the semilocal limit, the components decouple, and for the scalar component, the stability equation of semilocal strings is recovered. The dark sector affects the relevant sector of the perturbation equations through the appearance of the field $f_d$ in the scalar, and $x$ in the $W$ components, and through the deformation of the background solution in the functions $f$ and $\mathfrak{z}$.

The radial equations (\ref{eq:radeqEigM}) have been solved with the shooting to a fitting point method \cite{numrec}, as were the radial equations of the background vortex, Eq.\ (\ref{eq:radeq}). Our numerical methods were found to be stable for $M_{\scriptscriptstyle S} \sim M_{\scriptscriptstyle H}$.

The details of the calculations in this section are relegated to Appendix \ref{app:lin}.

\subsection{Domain of stability}\label{ssec:dstab}

As a validation of our code we have reproduced the domain of the stability of Z-strings in the Salam-Weinberg model (electroweak strings) and compared it to the data of Ref.\ \cite{GHelectroweak}. In our model, $\varepsilon=\theta_s=0$ corresponds to the case of the electroweak strings (with the dark sector decoupled).

Our method was as follows: we set $M_{\scriptscriptstyle Z}$, $M_{\scriptscriptstyle W}$, and $e$ to their physical values \cite{PDB}, and initially, $M_{\scriptscriptstyle H}$ as well, and $M_{\scriptscriptstyle S}^2 = M_{\scriptscriptstyle H}^2 \pm 2000\,{\rm GeV}^2$. Then we first lowered $M_{\scriptscriptstyle H}$ and $M_{\scriptscriptstyle S}$ keeping $M_{\scriptscriptstyle S}/M_{\scriptscriptstyle H}$ fixed, and then approached the semilocal limit, i.e., increased $\theta_{\scriptscriptstyle\rm W}$ towards $\pi/2$ while keeping $\bar{g}$, $\hat{g}$, $\varepsilon$, and the scalar potential parameters fixed, until $\Omega^2=0$ was reached (i.e., as long as there was a negative eigenvalue).

Our results for the case of no GKM are summarised in Table\ \ref{tab:dat}, with data from Ref.\ \cite{GHelectroweak} added for comparison\footnote{The data of Ref.\ \cite{GHelectroweak} has been reconstructed from its Fig.\ 1, using the data points in the postscript version of the figure in the {\tt arXiv.org} version of the paper, \href{https://arxiv.org/abs/hep-ph/9505357}{\tt hep-ph/9505357}, and transforming back to physical quantities from postscript coordinates, as the original data was not available any more.}. There is an excellent agreement between our data, and that of Ref.\ \cite{GHelectroweak}.

The stability of electroweak strings is restricted to $\beta_1 < 1$ (i.e., a Higgs mass smaller than the Z boson mass), and close to the semilocal limit, $\theta_{\scriptscriptstyle\rm W}\to \pi/2$.

\begin{figure}[h!]
 \noindent\hfil\includegraphics{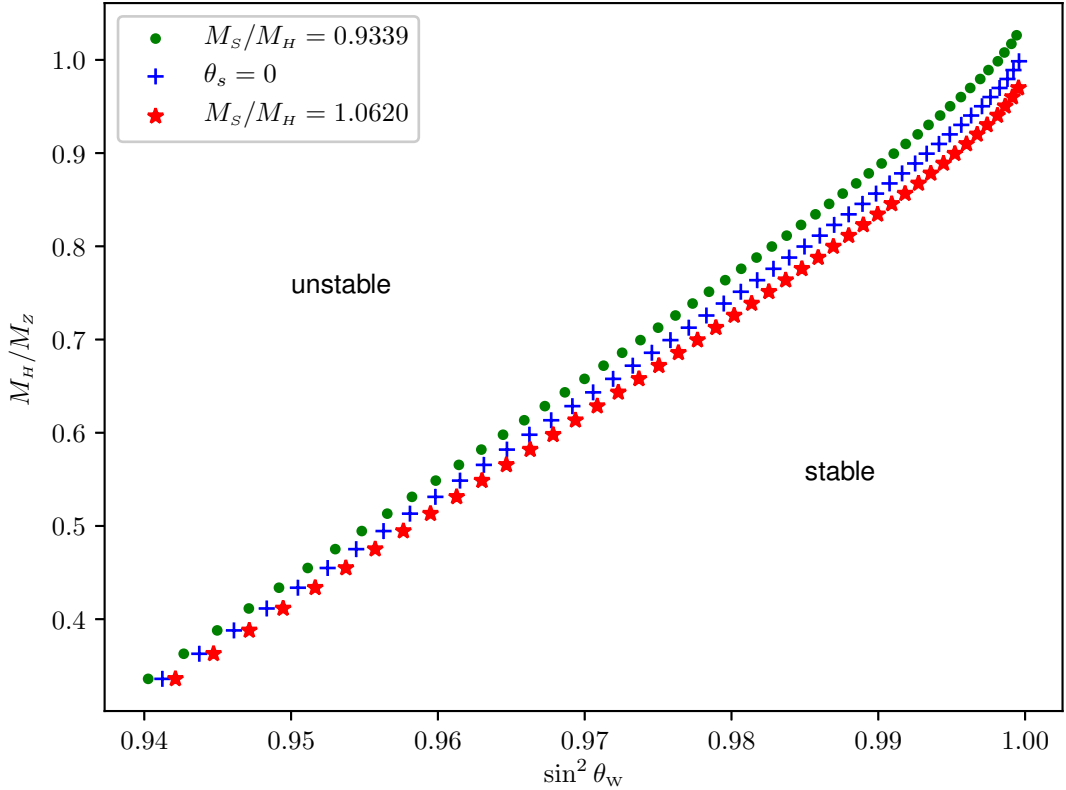}
 \caption{The boundary of the domain of stability, for
 $\varepsilon=0$, $\bar{g}=0.7416$, $\hat{g}=0.6172$, $\eta_1=173.4\,{\rm GeV}$, $\eta_2=217.4\,{\rm GeV}$, and $\theta_s=0.75$ compared to that of electroweak strings ($\theta_s=0$).
 %$\epsilon=0$ and $\theta_s=0.7$ and 0 (SM), and $\bar{g}=0.3638$,
 %$\eta_1 = 454.5\,{\rm GeV}$
 %$\eta_1=125655\,{\rm GeV}^2$
 %(physical values), $\hat{g}=1.57868$,
 %$\eta_2 = 424.9\,{\rm GeV}$.
 %$\eta_2^2=180561\,{\rm GeV}^2$.
 The domain of stability is as indicated on the figure.}
 \label{fig:npstab}
\end{figure}

In Fig.\ \ref{fig:npstab}, the effect of the Higgs portal coupling is shown. The motivation for this was the results for semilocal-dark strings in Ref.\ \cite{FLS}.
We have found that the Higgs portal coupling indeed has a stabilizing effect, however, in the experimentally undesirable parameter range, when the dark scalar is lighter than the Higgs. In the $M_{\scriptscriptstyle S} > M_{\scriptscriptstyle H}$ case, we actually found that adding the dark sector lowers the (already negative) eigenvalue, and narrows the domain of stability on, e.g.,  the $M_{\scriptscriptstyle H}/M_{\scriptscriptstyle Z}$ -- $\sin^2\theta_{\scriptscriptstyle\rm W}$ plane.

For an explanation, let us consider the potential for the perturbation function
$\delta \phi_1$, which is most relevant in the semilocal limit [see Eq.\ (\ref{eq:radeqPOM})],
\begin{equation}
 \label{eq:phi1pot}
 U=\beta_1 (f^2-1)+\beta' (f_d^2-\eta_{2s}^2) - g f^2/2\,,
\end{equation}
and estimate its value at the origin. Here $f(0)=0$, and we approximate the value of $f_d$ such that it minimises the potential $V$ of the theory when $f=0$, with $f_d^2 \approx \beta'/\beta_2 +\eta_{2s}^2$, yielding $U\approx -\beta_1 + (\beta')^2/\beta_2$. Expressing this with $\mu_{S,H}^2=M_{S,H}^2/(2g_{\scriptscriptstyle ZH}^2\eta_1^2)$ yields
\begin{equation}
 \label{eq:phi1potA}
 U(0)\approx \frac{2\mu_{\scriptscriptstyle H}^2[\mu_{\scriptscriptstyle H}^2(1-\cos2\theta_s)-\mu_{\scriptscriptstyle S}^2]}{\mu_{\scriptscriptstyle S}^2(1+\cos2\theta_s)-\mu_{\scriptscriptstyle H}^2(1-\cos2\theta_s)}\,,
\end{equation}
In the case of $M_{\scriptscriptstyle H} < M_{\scriptscriptstyle S}$, and $\theta_s$ close to $\pi/2$, this is a negative contribution.

\begin{figure}[b!]
 \noindent\hfil\includegraphics{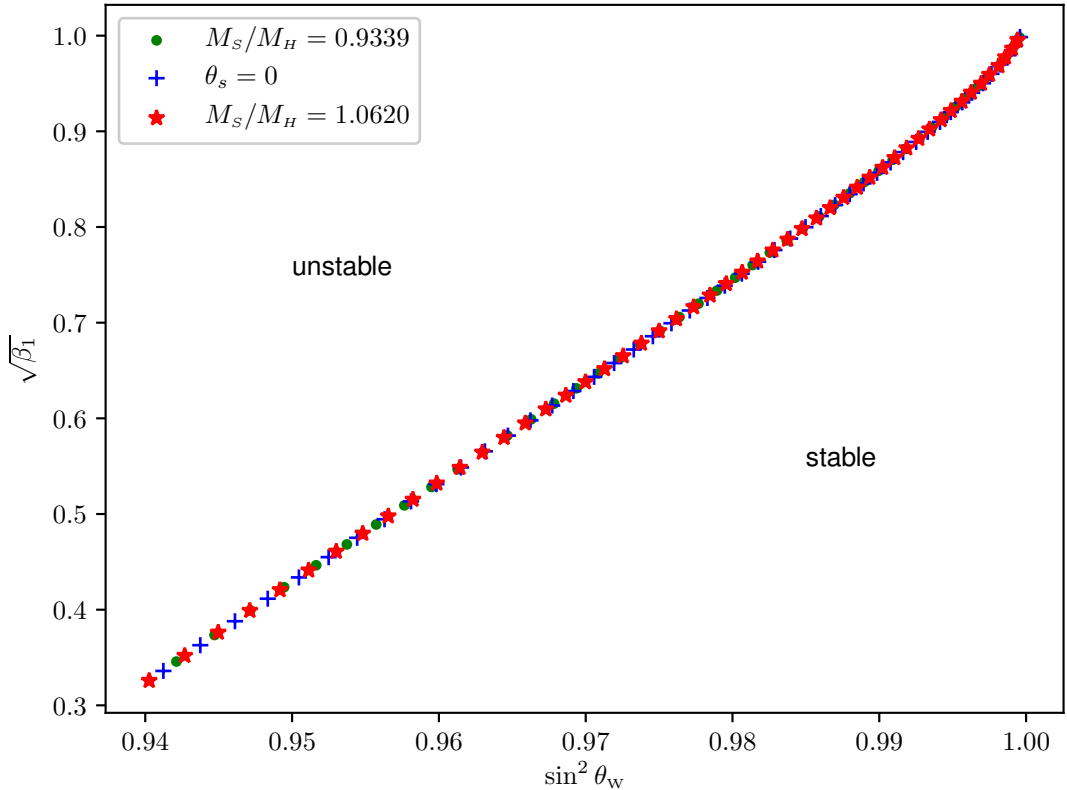}
\caption{Same as Fig.\ \ref{fig:npstab}, parametrised with $\sqrt{\beta_1}$ and $\sin^2\theta_{\scriptscriptstyle\rm W}$.}
 \label{fig:npstab2}
\end{figure}

\begin{figure}[b!]
 \noindent\hfil\includegraphics{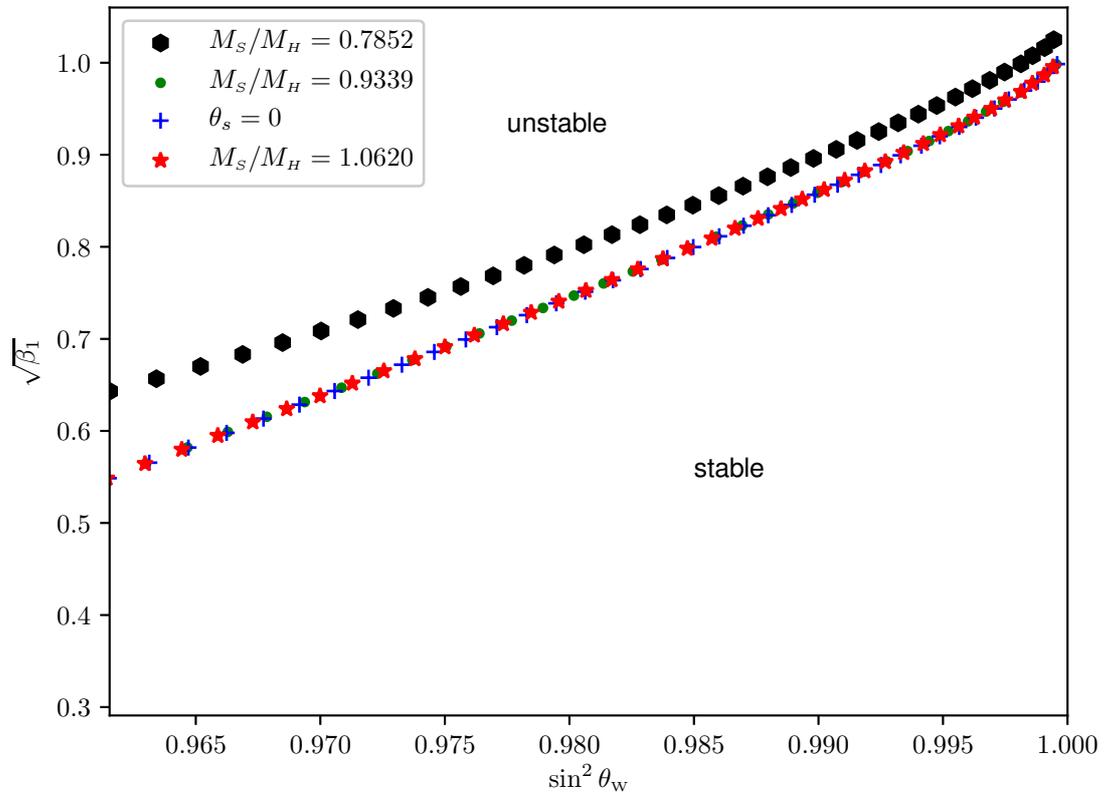}
\caption{Same as Fig.\ \ref{fig:npstab2}, with a lighter dark scalar.}
 \label{fig:npstab2l}
\end{figure}

It is found, that, quite remarkably, if the boundary curve of the domain of stability is plotted on the $\sqrt{\beta_1}$ -- $\sin^2\theta_{\scriptscriptstyle\rm W}$ plane (Fig.\ \ref{fig:npstab2}), the curves for different values of $M_{\scriptscriptstyle S}/M_{\scriptscriptstyle H}$ coincide. We have verified this coincidence numerically for $0.93 \le M_{\scriptscriptstyle S}/M_{\scriptscriptstyle H} \le 1.06$ and $0<\theta_s \le 0.75$. The differences between the value of $\sqrt{\beta_1}$ corresponding to the onset of instability between the cases considered is comparable to the numerical errors.
%(see Appendix \ref{app:coincide}).
The coincidence does not hold any more for $M_S/M_H = 0.7852$ (closer to
$M_S/M_H = 0.5$, where $h \to SS$ dark decays would contradict measurements; see Fig.\ \ref{fig:npstab2l}).
Because of this coincidence, in what follows, when we consider the effects of other parameters, and the Higgs and dark scalar masses are close enough, we shall only plot one curve in this parametrisation.

\begin{figure}
 \subfigure[{}]{
 \noindent\hfil\includegraphics[scale=.5]{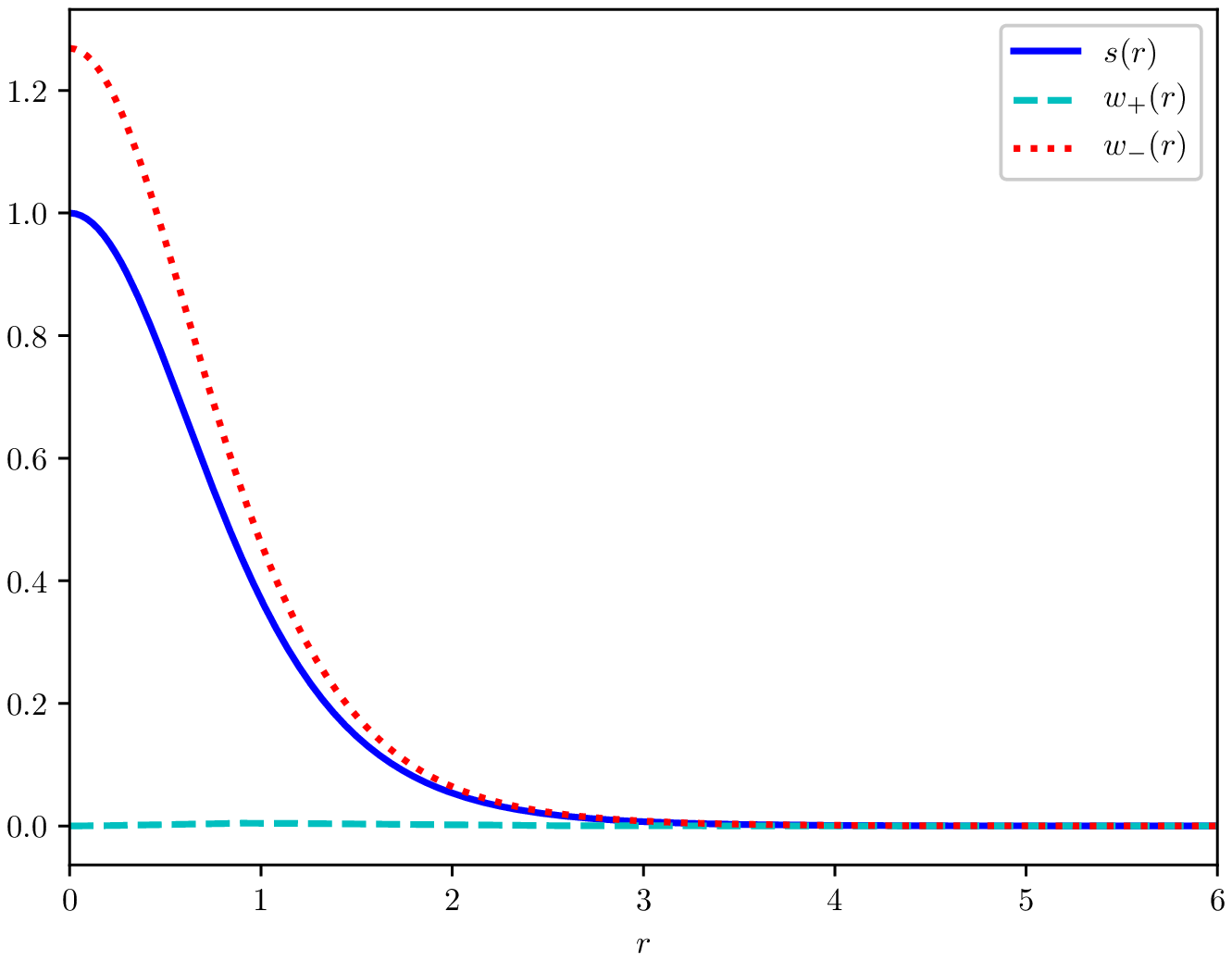}
 }
 \subfigure[{}]{
 \noindent\hfil\includegraphics[scale=.5]{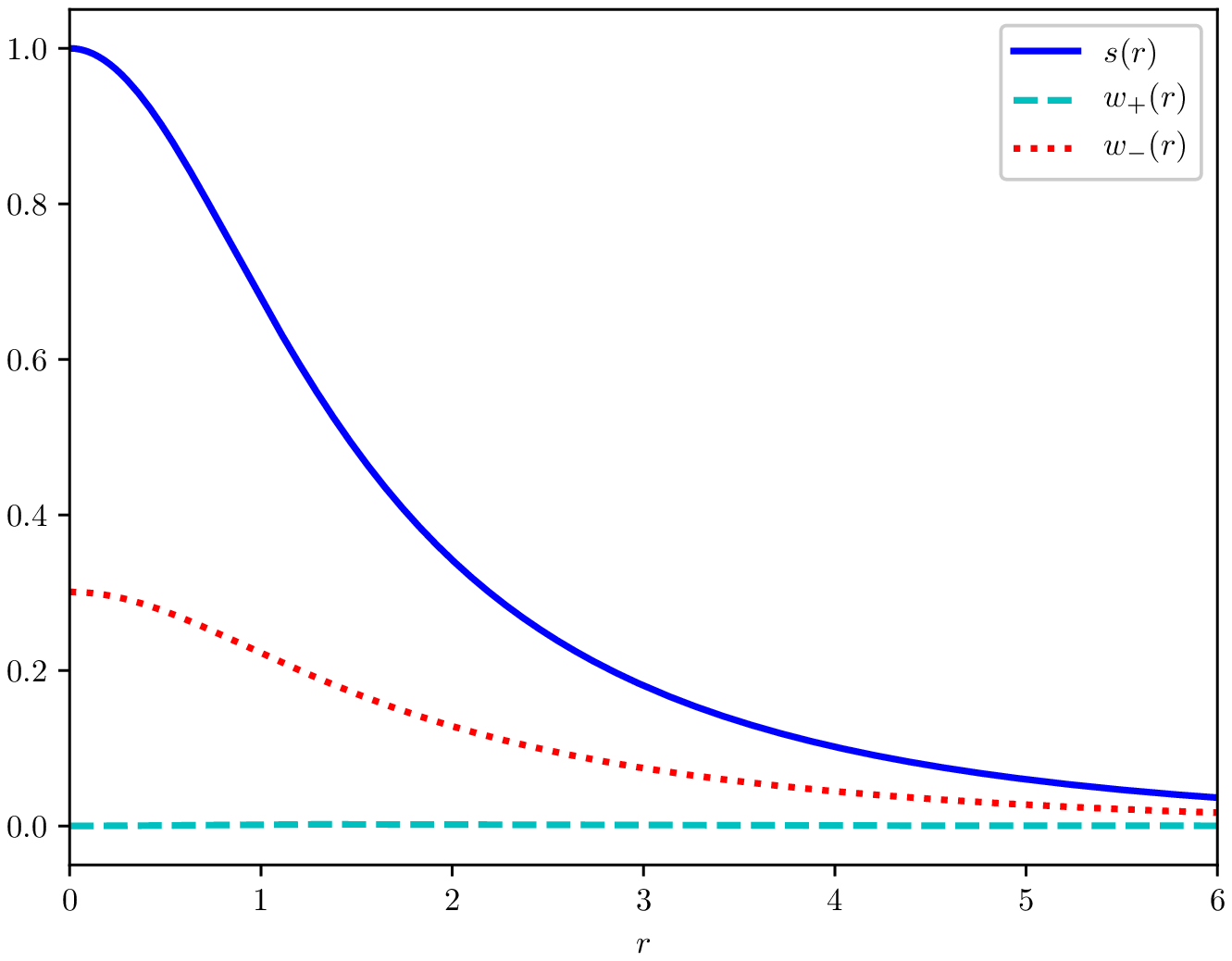}
 }
 \caption{The W and $\phi_1$ profile functions [$w_\pm(r)$ and $s(r)$, respectively] of the unstable eigenfunction (a) for physical parameters and (b) close to the semilocal limit. In both cases, $\theta_s=0$, $g_{\scriptscriptstyle X\phi^+}=0$, $M_{\scriptscriptstyle X}=94.87\,{\rm GeV}$, $M_{\scriptscriptstyle S}=132.8\,{\rm GeV}$.}
 \label{fig:eigenfunction}
\end{figure}

An explanation for the coincidence of the curves in Fig.\ \ref{fig:npstab2} is that the principal role in the instability is played by W condensation. This is the case for electroweak strings (see Refs.\ \cite{Perkins, GHelectroweak} and Fig.\ \ref{fig:eigenfunction}). The dark sector part of the background can be considered a perturbation for the allowed (small) values of the couplings between the visible and the dark sector. The allowed value of $\varepsilon$ is already rather small, and $\beta_2$ appears directly in the equation for the upper Higgs component, which is suppressed for $\theta_{\scriptscriptstyle\rm W} < \pi/2$: at the semilocal limit, $s_1(0)/w_-(0) \approx 3$ (and $w_+(0) \ll w_-(0)$), and at physical parameters $s_1(0)/w_-(0) \approx 0.8$,
which, in first order perturbation theory, would account for a suppression of the dark sector effects by a factor of $\sim 0.07$, which makes plausible both the coincidence of the curves in Fig.\ \ref{fig:npstab2}
and the suppression of the stabilisation effect upon leaving the semilocal limit.

%It is quite remarkable, that if we plot the boundary of the domain of stability on the $\sqrt{\beta_1}$ -- $\sin^2 \theta_{\scriptscriptstyle\rm W}$ plane (Fig.\ \ref{fig:npstab2}), the curves coincide. For the same value of $\sin^2\theta_{\scriptscriptstyle\rm W}$, the maximal values of $\sqrt{\beta_1}$ differ by $\sim 10^{-4}$.

In Table\ \ref{tab:dat}, we have collected some numerical data for reproducibility, and, for comparison, we have added the data points read off Fig.\ 1 of Ref.\ \cite{GHelectroweak}.

\begin{table}[h!]
\begin{center}
\begin{tabular}{c|c c c c}
% \hline
 $\sqrt\beta_1$ & $\sin^2\theta_{\scriptscriptstyle\rm W}$ &  & \\
%                & Ref.\ \cite{GHelectroweak} & SM     & $M_{\scriptscriptstyle S}^2 / M_{\scriptscriptstyle H}^2 = 0.8722$ &$M_{\scriptscriptstyle S}^2 / M_{\scriptscriptstyle H}^2 = 1.1279$\\
                & Ref.\ \cite{GHelectroweak} & electroweak    & $M_{\scriptscriptstyle S} / M_{\scriptscriptstyle H} = 0.9339$ &$M_{\scriptscriptstyle S} / M_{\scriptscriptstyle H} = 1.0620$\\
\hline
1               & 1.0                        & 0.9996 & 0.9995 & 0.9996\\
0.9             & 0.9910                     & 0.9933 & 0.9933 & 0.9933\\
0.8             & 0.9836                     & 0.9850 & 0.9849 & 0.9849\\
0.7             & 0.9756                     & 0.9758 & 0.9758 & 0.9758\\
0.6             & 0.9666                     & 0.9664 & 0.9664 & 0.9664\\
0.5             & 0.9576                     & 0.9568 & 0.9568 & 0.9568\\
0.4             & 0.9486                     & 0.9472 & 0.9472 & 0.9472\\
%\hline
\end{tabular}
\end{center}
\caption{Some points on the boundary of the domain of stability; for comparison, we also show data read off of Fig.\ 1 of Ref.\ \cite{GHelectroweak}. The other parameters are $\varepsilon=0$ and $\theta_s=0.75$ and 0 (electroweak), and $\bar{g}=0.7416$, $\eta_1=173.4\,{\rm GeV}$ (physical values), $\hat{g}=0.6172$, $\eta_2=217.4\,{\rm GeV}$.}
\label{tab:dat}
\end{table}

\begin{figure}[b!]
 \noindent\hfil\includegraphics{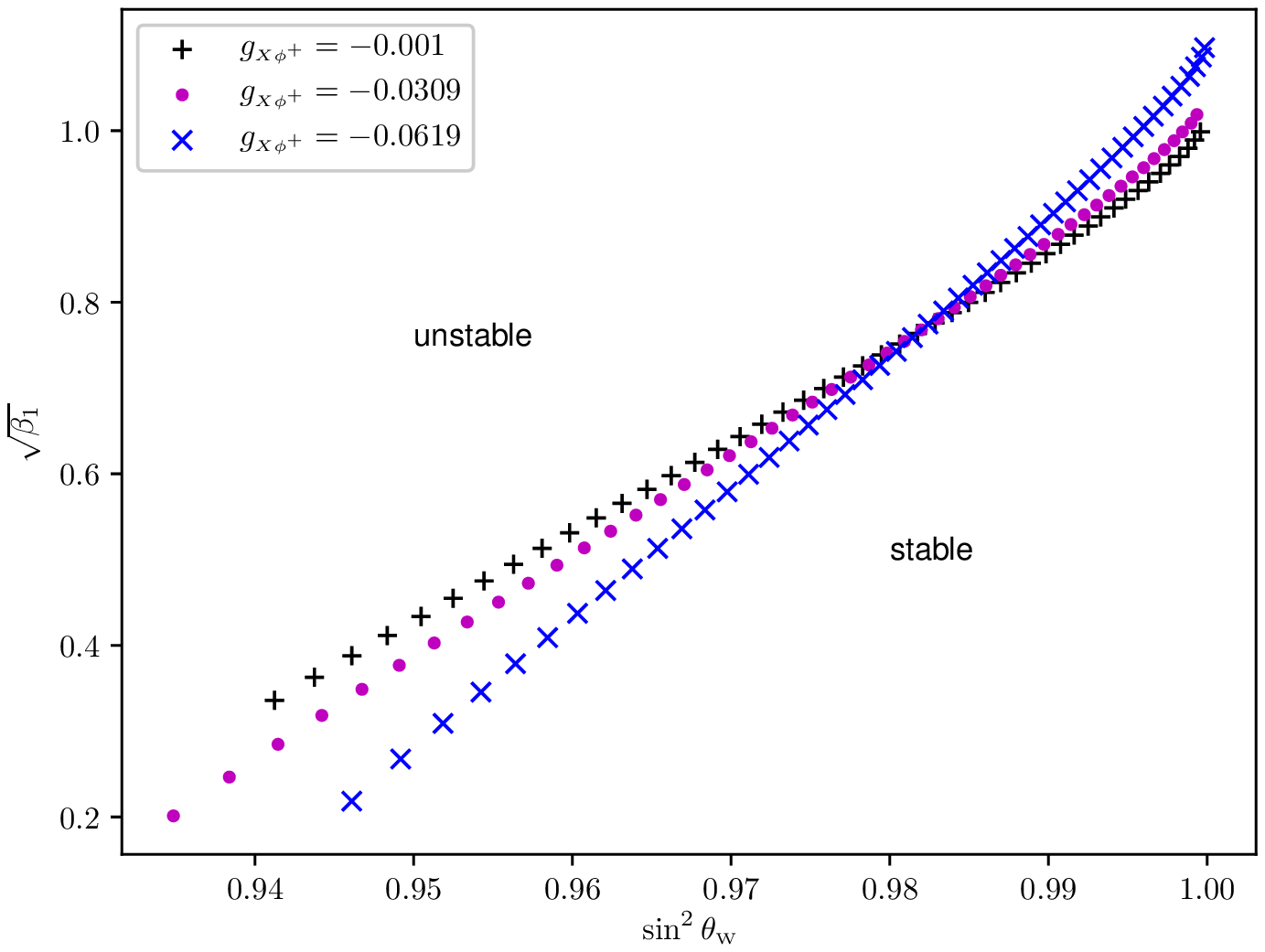}
\caption{The effect of gauge kinetic mixing on vortex stability. The  starting parameters ($M_{\scriptscriptstyle W}$, $M_{\scriptscriptstyle Z}$, $e$, $M_{\scriptscriptstyle H}$ physical and $g_{\scriptscriptstyle XS}=e$, $M_{\scriptscriptstyle S}^2=M_{\scriptscriptstyle H}^2+2000\,{\rm GeV}^2$, $\theta_s=0$, $M_{\scriptscriptstyle X}=94.87\,{\rm GeV}$ and $g_{\scriptscriptstyle X\phi^+}=-0.001$ and -0.0619) yield the parameters $\bar{g}=0.7416$, $\hat{g}=0.6172$, $\varepsilon=7.37\cdot10^{-5}$, $\eta_1=173.9\,{\rm GeV}$, $\eta_2=217.4\,{\rm GeV}$ and $\bar{g}=0.7362$, $\hat{g}=0.6406$, $\varepsilon=0.0446$, $\eta_1=175.7\,{\rm GeV}$, $\eta_2=208.6\,{\rm GeV}$.}
\label{fig:gkmeps}
\end{figure}

Another interaction, which is known to have a stabilising effect in the semilocal case is the GKM (see Ref.\ \cite{FLS}, where it is shown to lower the energy of semilocal-dark strings). Fig.\ \ref{fig:gkmeps} shows the effect of the GKM on the domain of stability. We have found that at the semilocal limit, the enhancement in the value of the quartic potential coefficient $\beta_1$ corresponding to zero eigenvalue (the upper edge of the domain of stability)  is significant for a large GKM; however, this is rapidly reduced by tuning $\theta_{\scriptscriptstyle\rm W}$ away from $\pi/2$. Also, experimental bounds do not allow the GKM to be large unless the dark gauge boson is heavy. For values of $\varepsilon$ consistent with experiment (Fig.\ \ref{fig:gkmeps} is for a value of $\varepsilon$ that is already at the limit), GKM results merely in an $\mathcal{O}(\varepsilon^2)$ correction.

\begin{figure}
 \noindent\hfil\includegraphics{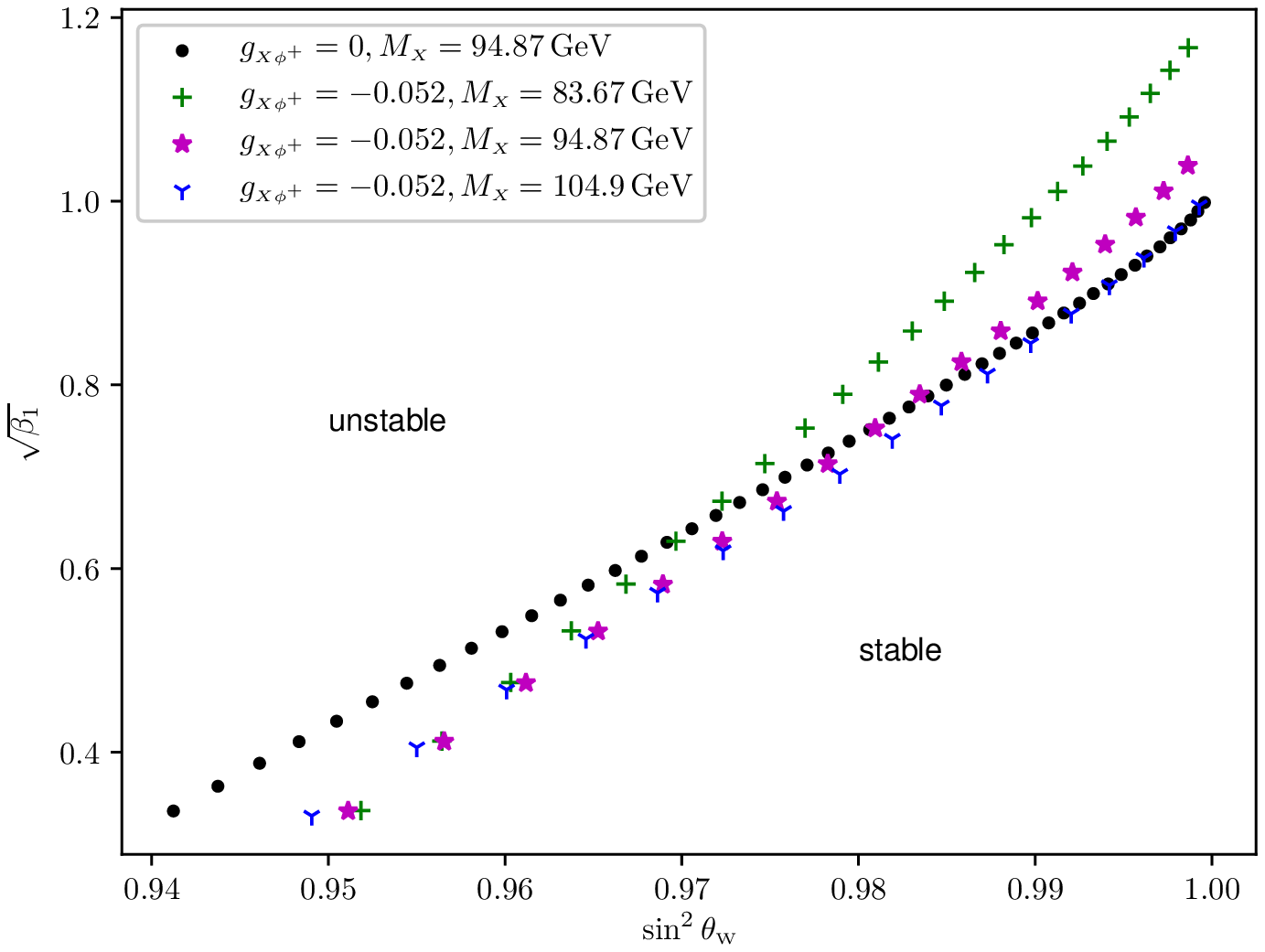}
\caption{The effect of different dark gauge boson masses on the stability in the case of large GKM. The starting parameters are $M_{\scriptscriptstyle W}$, $M_{\scriptscriptstyle Z}$, $e$, $M_{\scriptscriptstyle H}$ physical, $M_{\scriptscriptstyle S}^2=M_{\scriptscriptstyle H}^2+2000\,{\rm GeV}^2$, $\theta_s=0$, and $g_{\scriptscriptstyle X\phi^+}=0$ (electroweak), respectively, $g_{\scriptscriptstyle X\phi^+}=-0.052$ and different values of $M_{\scriptscriptstyle X}$.}
\label{fig:gkmeps2}
\end{figure}

\begin{figure}[t!]
\noindent\hfil\includegraphics{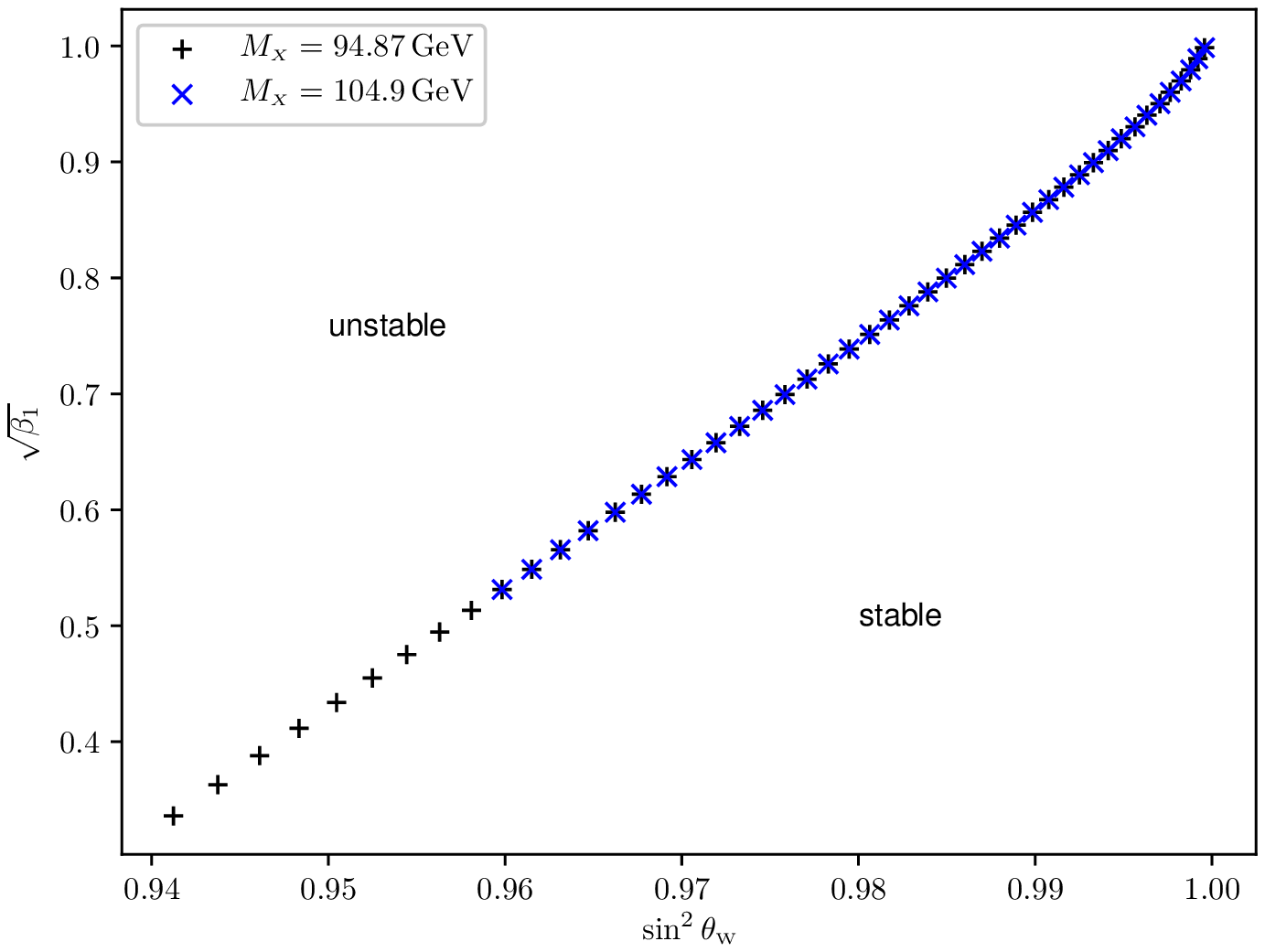}
\caption{The effect of the dark gauge boson mass in the case of no GKM, $g_{\scriptscriptstyle X\phi^+}=0$. The starting parameters are $M_{\scriptscriptstyle W}$, $M_{\scriptscriptstyle Z}$, $e$, $M_{\scriptscriptstyle H}$ physical, $M_{\scriptscriptstyle S}^2=M_{\scriptscriptstyle H}^2+2000\,{\rm GeV}^2$, $\theta_s=0$, and $g_{\scriptscriptstyle X\phi^+}=0$.}
 \label{fig:gkmepsMX}
 \end{figure}

In Fig.\ \ref{fig:gkmeps2}, the effect of the mass of the dark gauge boson is shown.
The sensitivity to the dark gauge boson mass is in contrast to the insensitivity in the case of stabilisation by the scalar potential (i.e., no GKM, Fig.\ \ref{fig:gkmepsMX}).

\begin{figure}[b!]
 \noindent\hfil\includegraphics{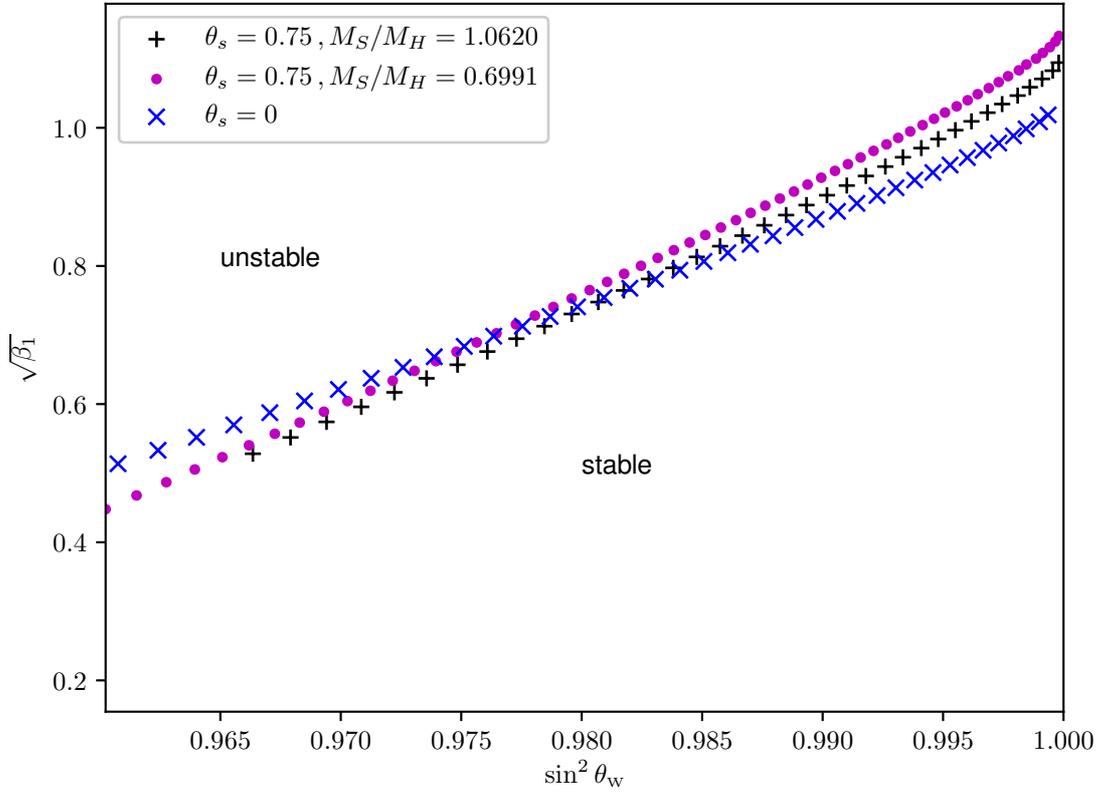}
\caption{The combined effects of the GKM and the scalar mixing; parameters as in Fig.\ \ref{fig:gkmeps}, $g_{X\phi^+}=0.0619$}
\label{fig:gkmeps3}
\end{figure}

In Fig.\ \ref{fig:gkmeps3}, the combined effect of the GKM $\varepsilon$ and the scalar mixing (for both the dark scalar lighter than the Higgs, and slightly heavier) is considered. The stabilising effect is still restricted close to the semilocal limit.

\subsection{The behaviour of the eigenvalue}

\begin{figure}[b!]
 \noindent\hfil\includegraphics{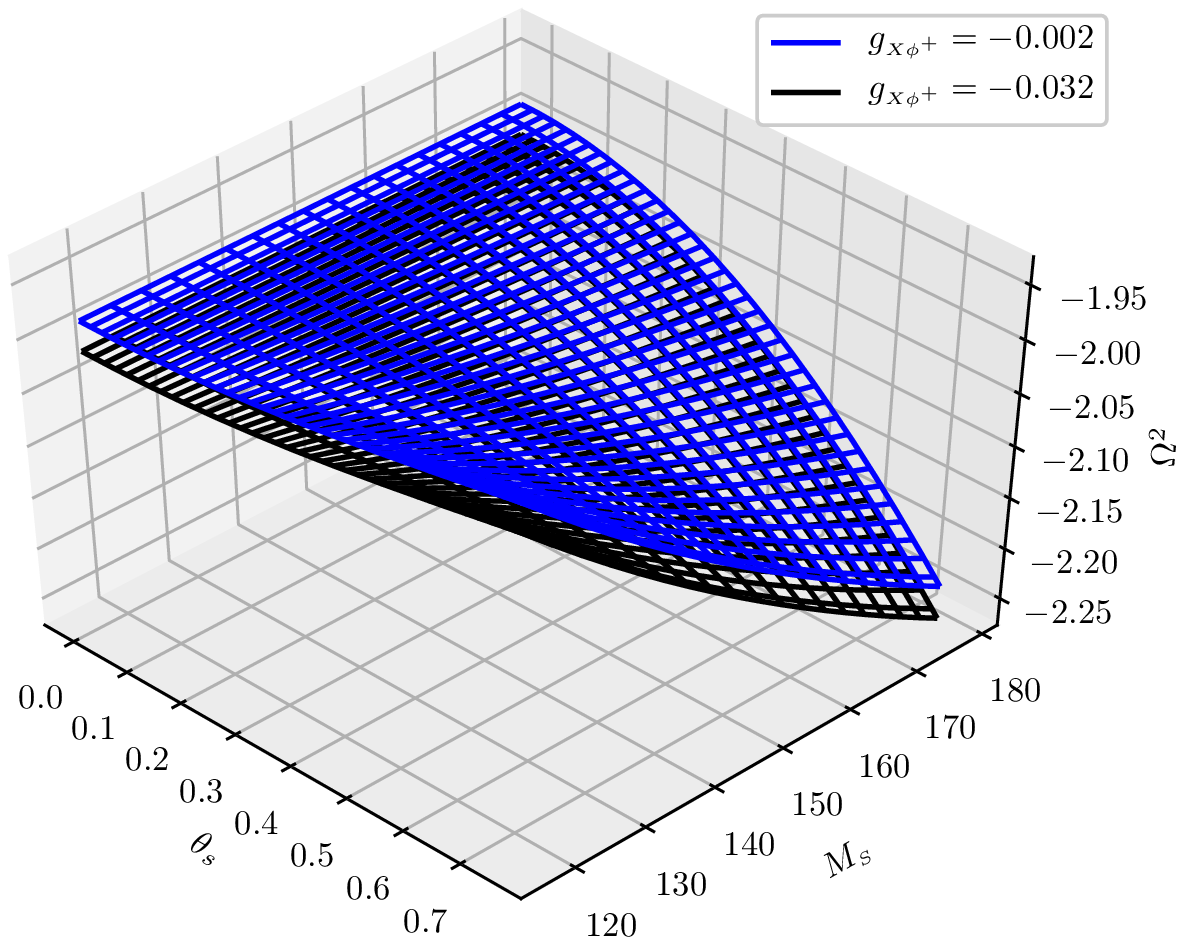}
 \caption{The eigenvalue of the stability equation (\ref{eq:radeqEigM}) as a function of $M_{\scriptscriptstyle S}$ and $\theta_s$, at $M_{\scriptscriptstyle Z}$, $M_{\scriptscriptstyle W}$, $e$ and $M_{\scriptscriptstyle H}$ physical, $M_{\scriptscriptstyle X}=94.87\,{\rm GeV}$, and $g_{\scriptscriptstyle XS}=e=0.3086$, $g_{\scriptscriptstyle X\phi^+}=-0.002$ and $-0.032$. %See also in colour online.
 }
 \label{fig:scan1}
\end{figure}

In order to assess the significance of the parameters, we have chosen a typical point, $M_{\scriptscriptstyle W}=80.4\,{\rm GeV}$, $M_{\scriptscriptstyle Z}=91.2\,{\rm GeV}$, $e=0.3086$, $M_{\scriptscriptstyle H}=125.1\,{\rm GeV}$ (physical values), $M_{\scriptscriptstyle X}=94.87\,{\rm GeV}$, $M_{\scriptscriptstyle S}=132.8\,{\rm GeV}$, $g_{\scriptscriptstyle X\phi^+}=0$, $g_{\scriptscriptstyle XS}=0.3086$ and $\theta_s=0.75$,
%$M_{\scriptscriptstyle X}=94.868\,{\rm GeV}$, $g_{\scriptscriptstyle XS}=e=0.15154$, $g_{\scriptscriptstyle X\phi^+}=-0.002$,
and obtained the derivatives of the eigenvalue with respect to the parameters. These are collected in Table\ \ref{tab:derivs}. We have concluded, that the parameters with the largest influence are $M_{\scriptscriptstyle S}$ and $\theta_s$.

\begin{table}
\begin{center}
\begin{tabular}{|c|c|}
\hline
Parameter & Derivative \\
\hline
\hline
$g_{\scriptscriptstyle X\phi^+}$ & 0 (parabolic maximum)\\
$M_{\scriptscriptstyle X}$        & $-9.02\cdot 10^{-8}\,{\rm GeV}^{-1}$\\
$g_{\scriptscriptstyle XS}$     & $2.77\cdot 10^{-5}$\\
$M_{\scriptscriptstyle S}$        & $-5.57\cdot 10^{-3}\,{\rm GeV}^{-1}$\\
$\theta_s$   & $-9.87\cdot 10^{-2}$\\
%$M_{\scriptscriptstyle S}^2$      & $-8.557\cdot 10^{-6}$\\
%$M_{\scriptscriptstyle X}^2$      & \ \ $1.313\cdot10^{-7}$\\
%$\theta_s$   & $-0.1656$\quad\quad\quad\\
%$g_{\scriptscriptstyle X\phi^+}$& $0.3373$\quad\ \ \\
%$g_{\scriptscriptstyle XS}$     & \ \ $2.653\cdot 10^{-4}$\\
\hline
\end{tabular}
\end{center}
\caption{Derivatives of the eigenvalue of the stability equation (\ref{eq:radeqEigM}) with respect to model parameters at $M_{\scriptscriptstyle W}=80.4\,{\rm GeV}$, $M_{\scriptscriptstyle Z}=91.2\,{\rm GeV}$, $e=0.3086$, $M_{\scriptscriptstyle H}=125.1\,{\rm GeV}$ (physical values), $M_{\scriptscriptstyle X}=94.87\,{\rm GeV}$, $M_{\scriptscriptstyle S}=132.8\,{\rm GeV}$, $g_{\scriptscriptstyle X\phi^+}=0$, $g_{\scriptscriptstyle XS}=0.3086$ and $\theta_s=0.75$. Note that $-\Omega^2$ is the squared growth rate corresponding to rescaled time, or, equivalently, $|g_{\scriptscriptstyle ZH}\eta_1 \Omega|$ is a growth rate in unscaled time. %or time measured in $1/(g_{\scriptscriptstyle ZH}\eta_1)$.
%, i.e., time in units of $1/(g_{\scriptscriptstyle ZH}\eta_1)^2$. The units of the derivatives are this ($1/{\rm GeV}^2$) divided by the units of the parameters.
Here %$g_{\scriptscriptstyle ZH}=-0.3708$,
$|g_{\scriptscriptstyle ZH}|\eta_1=64.49\,{\rm GeV}$.
%$1/(g_{\scriptscriptstyle ZH}\eta_1)=0.01551\,{\rm GeV}^{-1}$.%$1/(g_{\scriptscriptstyle ZH}\eta_1)^2=2.405\cdot10^{-4}\,{\rm GeV}^{-2}$.
}
\label{tab:derivs}
\end{table}

We have next varied $M_{\scriptscriptstyle S} > M_{\scriptscriptstyle H}$ (so that dark Higgs decays do not exclude the considered parameter values) and $\theta_s$, in the range $M_{\scriptscriptstyle H}^2 \le M_{\scriptscriptstyle S}^2 \le 2 M_{\scriptscriptstyle H}^2$ and $0\le \theta_s \le 1.5$. We have found no stable solutions. The eigenvalue seems to depend most strongly on the parameters $M_{\scriptscriptstyle S}^2$ and $\theta_s$.

In Fig.\ \ref{fig:scan1} we present numerical data of the eigenvalue $\Omega^2$ as a function of the two parameters that seem most relevant (i.e., they parametrise the scalar sector most directly), $M_{\scriptscriptstyle S}$ and $\theta_s$. Note, that the eigenvalue is always negative (signalling instability), and becomes more negative with larger values of the dark scalar mass $M_{\scriptscriptstyle S}$.

\begin{figure}[t!]
  \noindent\hfil\includegraphics{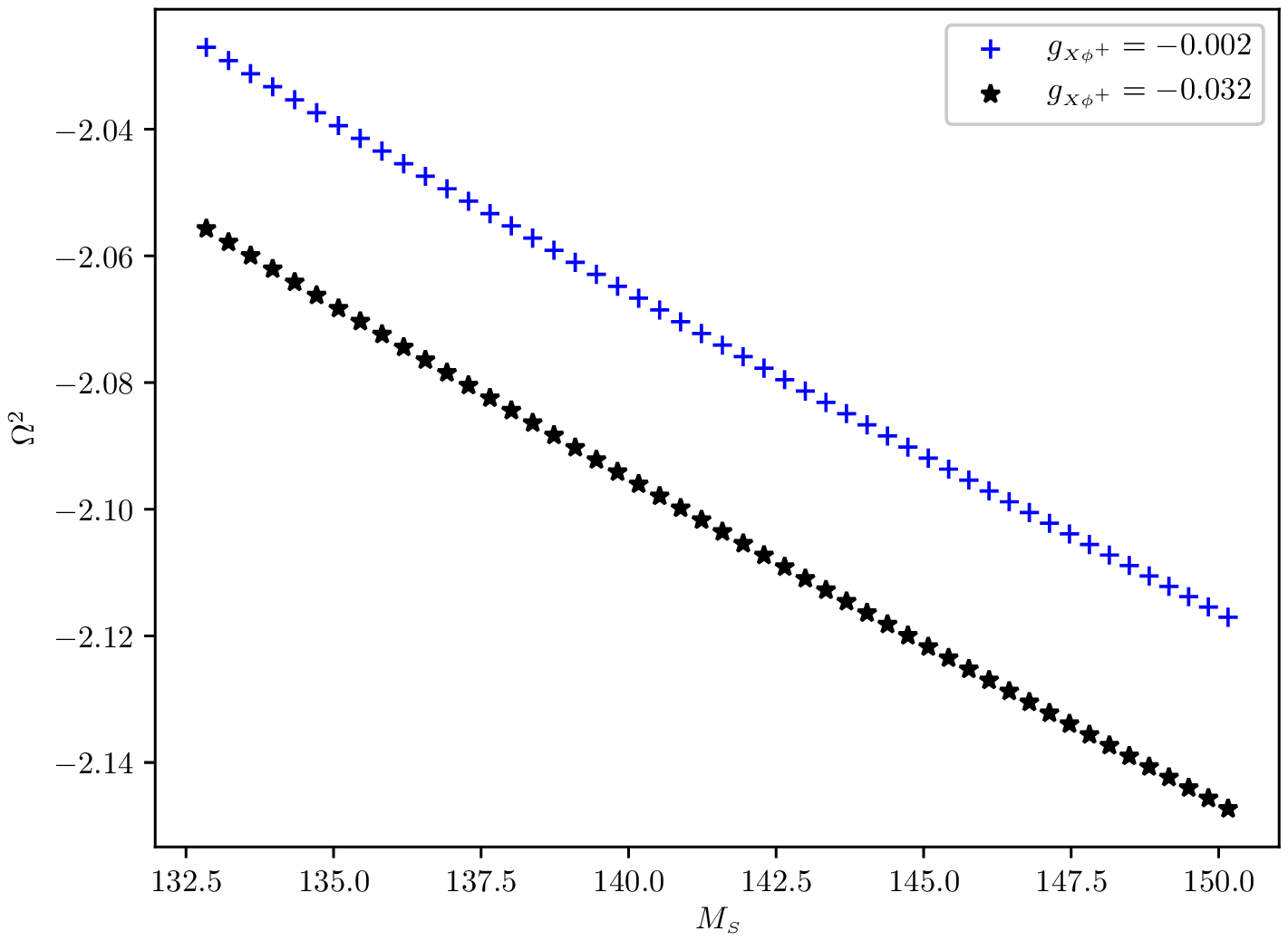}
  \caption{Typical $\Omega^2$--$M_{\scriptscriptstyle S}$ curves at
  $M_{\scriptscriptstyle Z}$, $M_{\scriptscriptstyle W}$, $e$ and $M_{\scriptscriptstyle H}$ physical, $M_{\scriptscriptstyle X}=94.868\,{\rm GeV}$, and $g_{\scriptscriptstyle XS}=e=0.3086$, $\theta_s=0.75$, $g_{\scriptscriptstyle X\phi^+}=-0.002$, and $-0.032$.
%
%  $g_{\scriptscriptstyle XS}=e=0.15154$, $g_{\scriptscriptstyle X\phi^+}=-0.032$ and $g_{\scriptscriptstyle X\phi^+}=-0.002$, and  $\theta_s=0.75$, $M_{\scriptscriptstyle X}=94.868\,{\rm GeV}$.
}
  \label{fig:Om2MS}
\end{figure}

\begin{figure}[h!]
 \noindent\hfil\includegraphics{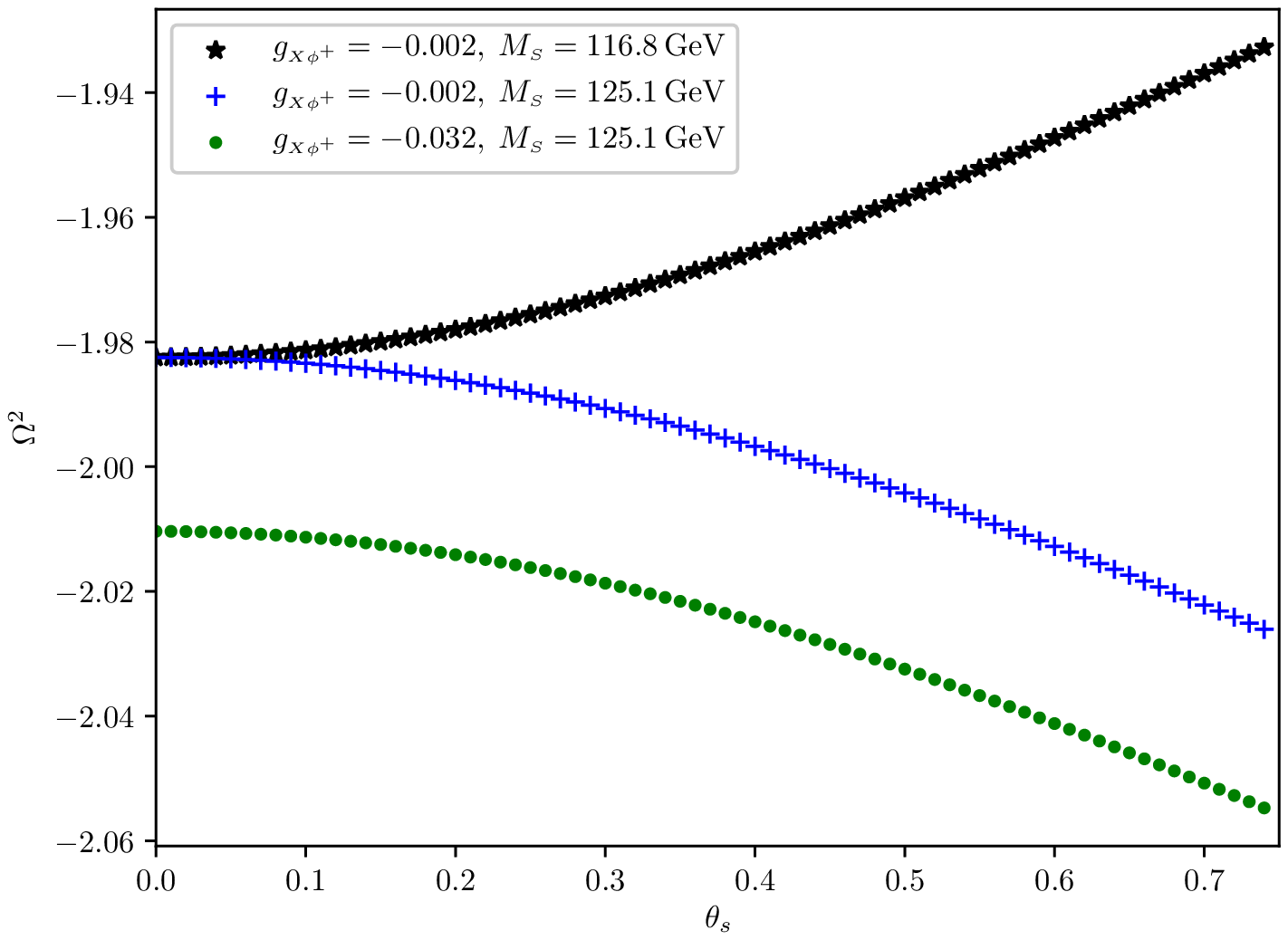}
 \caption{A typical $\Omega^2$--$\theta_s$ curves at $M_{\scriptscriptstyle Z}$, $M_{\scriptscriptstyle W}$, $e$ and $M_{\scriptscriptstyle H}$ physical, $M_{\scriptscriptstyle X}=94.868\,{\rm GeV}$, and $g_{\scriptscriptstyle XS}=e=0.3086$, $\theta_s=0.75$, $g_{\scriptscriptstyle X\phi^+}=-0.002$ and $-0.032$.
 %. $g_{\scriptscriptstyle XS}=e=0.15154$, $g_{\scriptscriptstyle X\phi^+}=-0.032$ and $g_{\scriptscriptstyle X\phi^+}=-0.002$, and $M_{\scriptscriptstyle X}=94.868\,{\rm GeV}$.
 }
\label{fig:Om2thetas}
\end{figure}

In Fig.\ \ref{fig:Om2MS} a typical $\Omega^2$ -- $M_{\scriptscriptstyle S}$ curve is shown for
$M_{\scriptscriptstyle Z}$, $M_{\scriptscriptstyle W}$, $e$ and $M_{\scriptscriptstyle H}$ physical, $M_{\scriptscriptstyle X}=94.868\,{\rm GeV}$, and $g_{\scriptscriptstyle XS}=e=0.3086$, $\theta_s=0.75$, $g_{\scriptscriptstyle X\phi^+}=-0.002$, and $-0.032$.
%$g_{\scriptscriptstyle XS}=e=0.15154$, $g_{\scriptscriptstyle X\phi^+}=-0.032$, $\theta_s=0.75$.
The eigenvalues are clearly negative and descending as a function of $M_{\scriptscriptstyle S}$. Similarly, Fig.\ \ref{fig:Om2thetas} shows a typical $\Omega^2$ -- $\theta_s$ curve in the parameter range studied.
The curves in Fig.\ \ref{fig:Om2MS} and  in Fig.\ \ref{fig:Om2thetas} are cross sections of the surfaces in Fig.\ \ref{fig:scan1}. In Fig.\ \ref{fig:Om2thetas}, we have added an additional curve for $M_{\scriptscriptstyle S} < M_{\scriptscriptstyle H}$. One interesting feature of Fig.\ \ref{fig:Om2thetas} is that the eigenvalue has a minimum for $M_S > M_H$ (and maximum for $M_S < M_H$) at $\theta_s=0$ (and thus also for $\beta'=0$, $\lambda'=0$), i.e., for small values of the GKM its sign is not important.

The data indicate clearly that in the physically relevant parameter range where the dark gauge boson mass is $M_X \gtrsim M_Z$, the scalar mass is $M_H/2 < M_S < M_H$, the scalar mixing angle is $|\theta_s| \lesssim 1$, and the dark charge is $g_{XS}\sim e$, no stable solutions exist. In this parameter range, we have found, that a larger dark scalar mass corresponds to stronger instability. %, which, however, may not be valid for a different parameter regime.
On the other hand, for $M_{\scriptscriptstyle S} < M_{\scriptscriptstyle H}$, the addition of the dark sector (non-zero scalar mixing angle $\theta_s$, and, similarly, GKM $\varepsilon$) results in stability properties of electroweak strings (although still not reaching the physical parameter values) that are significantly better.

The fact that the stabilising effects are rather strong in the semilocal limit, %$\theta_{\scriptscriptstyle\rm W}\to \pi/2$,
and much weaker for smaller values of the Weinberg angle, is explained by the nature of the instability. In the semilocal model, the instability is due to the possibility of unwinding in the scalar ($\Phi$) sector \cite{hin1, hin2}; however, in the full non-Abelian theory, the instability also involves the condensation of $W$ bosons in the string core \cite{JPV1, JPV2, Perkins, GHelectroweak} (see also Fig.\ \ref{fig:eigenfunction}). In the present model, the dark sector only couples to the Higgs scalar and the weak hypercharge $U(1)$ fields, and the dark part of the background vortex does not influence the $W$ fields other than slightly distorting the visible sector part of the background.

\section{Summary and outlook}\label{sec:discussion}
In this paper, we have presented a study of electroweak-dark strings, complementing those of dark strings in Higgs portal models\cite{VachaspatiDS, Vachaspati1,Vachaspati2, Vachaspati3, HartmannArbabzadah, BrihayeHartmann}. We have demonstrated the main properties of the equations describing these strings, and their numerical solutions. We have shown that these strings exists at the well-known scale of electroweak strings, in contrast to the unknown scale of dark strings.

We have also examined the stability of the electroweak-dark strings. Close to the semilocal limit, we have demonstrated that the stabilising effect of the dark sector found in the semilocal case in Ref.\ \cite{FLS}, persists in the electroweak-dark case, i.e., the stability of electroweak strings is enhanced to $M_{\scriptscriptstyle H}/M_{\scriptscriptstyle Z} >1$; however, this happens for values of the parameters of the model excluded by experimental bounds: for large gauge-kinetic mixing $\varepsilon$ with the light dark sector Abelian gauge boson ($M_{\scriptscriptstyle X} \lesssim M_{\scriptscriptstyle Z}$) or large scalar mixing (Higgs portal coupling) and light dark scalar $M_{\scriptscriptstyle S} < M_{\scriptscriptstyle H}$. For $M_{\scriptscriptstyle S} > M_{\scriptscriptstyle H}$, we have considered the parameter range experimentally allowed and found instabilities. Complemented with the fact that in the limit $M_{\scriptscriptstyle S} \to \infty$, the instabilities in the electroweak case are recovered, one can conclude that in the model considered, there is no stabilisation due to the interaction with dark sector fields.

The properties of the eigenfunction of the linearised equation corresponding to the instability sheds light on the reasons why the stabilising effects do not persist to lower values of the Weinberg angle. For those values, the components corresponding to the $W$ fields are large, the mechanism of the instability is W condensation, and the couplings considered here affect primarily the Higgs and the Z fields.

In future studies, the analysis may be supplemented by considering fermionic fields. In the electroweak case, the topological consequences of fermionic zero modes have been considered in Refs.\ \cite{liuvach, grovesperkins, stojk1, stojk2, stojk3}, suggesting that an interplay of fermionic modes and the deformations corresponding to the unstable modes results in new, stable electroweak strings. The effect of the Dirac sea has also been considered for electroweak strings; Ref.\ \cite{Naculich} finds instabilities due to light fields, which may be stabilised by filled fermionic states, whereas Refs.\ \cite{Weigel1, Weigel2} find stabilisation due to heavy fermions. Another line of research may be the consideration of models with couplings to the W fields. It should be emphasised, however, that LEP electroweak measurements put stringent bounds on not-too-heavy fields coupled to the electroweak model.

\paragraph{Acknowledgement} We acknowledge the support of  the Spanish Ministerio de Ciencia, Innovación y Universidades (Grant No.\ PCI2018-092896) and the EU (QuantERA CEBBEC).

We thank the Referee for useful suggestions.

\appendix
\section{Details of the calculations}\label{app:details}
To obtain the solutions and assess their stability, we start with the field equations derived from the gauge and scalar Lagrangians (\ref{eq:LewG2}) and (\ref{eq:Ls}),
\begin{equation}
 \label{eq:FieldEq}
 \begin{aligned}
  D_\mu D^\mu \Phi &= -2\lambda_1(\Phi^\dagger\Phi-\eta_1^2)\Phi-\lambda'(\chi^*\chi-\eta_2^2)\Phi\,,\\
  \tilde{D}_\mu{\tilde{D}}^\mu \chi &= -2\lambda_2(\chi^*\chi-\eta_2^2)\chi -\lambda'(\Phi^\dagger\Phi-\eta_1^2)\chi\,,\\
  \partial_\mu F^{\mu\nu} &= J^\nu_{\rm el}+J^\nu_{\rm el,g}-g\alpha_1\partial_\mu(W^{\mu1}W^{\nu2}-W^{\nu1}W^{\mu2})\,,\\
  \partial_\mu Z^{\mu\nu} &= J^\nu_{\scriptscriptstyle Z}+J^\nu_{{\scriptscriptstyle Z},g}-g\alpha_2\partial_\mu(W^{\mu1}W^{\nu2}-W^{\nu1}W^{\mu2})\,,\\
  \partial_\mu X^{\mu\nu} &= J^\nu_{\scriptscriptstyle X}+J^\nu_{{\scriptscriptstyle X},g}-g\alpha_3\partial_\mu(W^{\mu1}W^{\nu2}-W^{\nu1}W^{\mu2})\,,\\
  \partial_\mu W^{\mu\nu a} &= J^{\nu a}_{\scriptscriptstyle W} + J^{\nu a}_{{\scriptscriptstyle W},g}\,,
 \end{aligned}
\end{equation}
where $W^{\mu\nu a}=\tilde{W}^{\mu\nu a}+g\varepsilon^{ab}(W^{\mu b}W^{\nu 3}-W^{\nu b}W^{\mu 3})$, $\varepsilon^{ab}$ is antisymmetric and $\varepsilon^{12}=1$, the Abelian currents are given by
\begin{equation}
 \label{eq:currentsA}
 \begin{aligned}
  J^\nu_{\rm el} &= \imagi g_{\scriptscriptstyle A\phi 1}((D^\nu\Phi)_1^\dagger \phi_1 -\phi_1^*(D^\nu\Phi)_1)\,,\\
  J^\nu_{\rm el,g} &= g\alpha_1 {\tilde W}^{\mu\nu 1}W_\mu^2 - g\alpha_1 {\tilde W}^{\mu\nu 2}W_\mu^1 +g^2 \alpha_1 W_\mu^a W^{\mu a}W^{\nu 3}-g^2 \alpha_1 W_\mu^3 W^{\mu a}W^{\nu a}\,,\\
  J^\nu_{\scriptscriptstyle Z} &= \imagi g_{\scriptscriptstyle Z\phi a}((D^\nu\Phi)_a^\dagger \phi_a -\phi_a^*(D^\nu\Phi)_a)
  +\imagi g_{\scriptscriptstyle ZS}(({\tilde D}^\nu\chi)^* \chi -\chi^*({\tilde D}^\nu\chi))\,,\\
  J^\nu_{{\scriptscriptstyle Z},g} &= g\alpha_2 {\tilde W}^{\mu\nu 1}W_\mu^2 - g\alpha_2 {\tilde W}^{\mu\nu 2}W_\mu^1 +g^2 \alpha_2 W_\mu^a W^{\mu a}W^{\nu 3}-g^2 \alpha_2 W_\mu^3 W^{\mu a}W^{\nu a}\,,\\
  J^\nu_{\scriptscriptstyle  X} &= \imagi g_{\scriptscriptstyle X\phi a}((D^\nu\Phi)_a^\dagger \phi_a -\phi_a^*(D^\nu\Phi)_a)
  +\imagi g_{\scriptscriptstyle XS}(({\tilde D}^\nu\chi)^* \chi -\chi^*({\tilde D}^\nu\chi))\,,\\
  J^\nu_{{\scriptscriptstyle X},g} &= g\alpha_3 {\tilde W}^{\mu\nu 1}W_\mu^2 - g\alpha_3 {\tilde W}^{\mu\nu 2}W_\mu^1 +g^2 \alpha_3 W_\mu^a W^{\mu a}W^{\nu 3}-g^2 \alpha_3 W_\mu^3 W^{\mu a}W^{\nu a}\,,\\
 \end{aligned}
\end{equation}
and the non-Abelian one as
\begin{equation}
 \label{eq:currentsnA}
 \begin{aligned}
  J_{\scriptscriptstyle  W}^{\nu a} &= \frac{\imagi g}{2}( D^\nu\Phi^\dagger \tau^a \Phi - \Phi^\dagger \tau^a D^\nu \Phi)\,,\\
  J_{{\scriptscriptstyle W},g}^{\nu a} &= -g {\tilde W}^{\mu \nu 3}\varepsilon^{ab}W_\mu^b - g \varepsilon^{ab}{\tilde W}^{\mu\nu b}W_\mu^3 \\
  &\ \ -g^2 W_\mu^b W^{\mu b} W^{\nu a}+g^2 W_\mu^3 W^{\mu 3}W^{\nu a}-g^2 W_\mu^b W^{\nu b}W^{\mu a}-g^2 W^{3 \nu}W_\mu^3 W^{\mu a}\,,
 \end{aligned}
\end{equation}
where $\tau^a$ denote the Pauli matrices. In Eqs.\ (\ref{eq:currentsA}) and (\ref{eq:currentsnA}), $a=1,2$, and the notation $g_{\scriptscriptstyle ZH}=g_{\scriptscriptstyle X\phi_2}$,
$g_{\scriptscriptstyle XH}=g_{\scriptscriptstyle X\phi_2}$, $g_{\scriptscriptstyle Z\phi^+}=g_{\scriptscriptstyle X\phi_1}$, and $g_{\scriptscriptstyle X\phi^+}=g_{\scriptscriptstyle X\phi_1}$ is used (see Ref.\ \cite{Vachaspati1}).

\subsection{Linearised equations}\label{app:lin}
Let us add perturbations to the vortex solution, $A_\mu \to \delta A_\mu$, $W_\mu^a\to\delta W_\mu^a$, $Z_\mu\to Z_\mu+\delta Z_\mu$, $X_\mu\to X_\mu+\delta X_\mu$, $\phi_a\to \phi_a+\delta\phi_a$ and $\chi
\to\chi+\delta\chi$.
In the analysis of vortex perturbations, we follow the lines of Refs.\ \cite{Goodband, GHelectroweak}; see also Refs.\ \cite{FL, FLS, twistedinstab1, twistedinstab2}.

To obtain simple linear equations, a gauge choice is of utmost importance. In the Abelian sector, we shall use the background field gauge of Refs.\ \cite{baacke, Goodband}, whereas for the non-Abelian gauge fields, we prescribe the background field gauge used in Ref.\ \cite{GHelectroweak}. This gauge choice, shown in Eq.\ (\ref{eq:gaugechoiceM}),
%\begin{equation}
% \label{eq:gaugechoice}
% \begin{aligned}
%  F_1 &= \partial_\mu \delta W^{\mu+} -\imagi g W_\mu^3 \delta W^{\mu+}-\frac{\imagi g}{\sqrt{2}}\phi_2^*\delta\phi_1=0\,,\\
%  F_2 &= \partial_\mu \delta W^{\mu-} +\imagi g W_\mu^3 \delta W^{\mu-}+\frac{\imagi g}{\sqrt{2}}\phi_2\delta\phi_1^*=0\,,\\
%  F_3 &= \partial_\mu\delta Z^\mu + \imagi g_{\scriptscriptstyle ZH}(\phi_2 \delta\phi_2^*-\phi_2^* \delta\phi_2)
%  +\imagi g_{\scriptscriptstyle ZS}(\chi \delta\chi^*-\chi^* \delta\chi_2)=0\,,\\
%  F_4 &= \partial_\mu\delta X^\mu + \imagi g_{\scriptscriptstyle XH}(\phi_2 \delta\phi_2^*-\phi_2^* \delta\phi_2)
%  +\imagi g_{\scriptscriptstyle XS}(\chi \delta\chi^*-\chi^* \delta\chi_2)=0\,.
% \end{aligned}
%\end{equation}
%Note, that $F_2=F_1^*$. The gauge fixing (\ref{eq:gaugechoice})
cancels linear first order derivatives of the gauge field perturbations, and in this way makes the separation of time derivatives possible. Note, that $F_2=F_1^*$.

Let $\Psi=(\delta A_\mu, \delta W^+_\mu,\delta W^-_\mu, \delta Z_\mu, \delta X_\mu, \delta \phi_a, \delta\phi_a^*, \delta\chi, \delta\chi^*)$ denote the components of the linear perturbations added to the fields.

Because of the background solution possessing a global direction in internal space (i.e., $\phi_1=0$) in the gauge used for the Ansatz (\ref{eq:Ansatz1}), (\ref{eq:Ansatz2}), the equations separate into decoupled blocks: (i) $A_\mu$, (ii) $\delta W^+_\mu$ and $\delta\phi_1$, (iii) $\delta W^-_\mu$ and $\delta \phi_1^*$, and (iv) $\delta Z_\mu$, $\delta X_\mu$, $\delta\phi_2$, $\delta\phi_2^*$, $\delta\chi$, and $\delta\chi^*$, in each block satisfying an equation of the form
\begin{equation}\label{eq:lineq}
 \mathcal{D}^I \Psi^I = 0\,,\quad I={\rm i},\dots,{\rm iv}\,,
\end{equation}
where $\mathcal{D}^I$ is a matrix with differential operators in the diagonal and coupling terms in the remaining elements.

The field $A_\mu$ is completely decoupled, $\Psi^{{\rm i}}_\mu =\delta A_\mu$, and
\begin{equation}\label{eq:operDi}
 \mathcal{D}^{{\rm i}} = \square\,,
\end{equation}
diagonal in the Lorentz index. As the electromagnetic field is decoupled, it does not influence the stability of the string. In what follows, it is not considered further.

In block (ii), the fields are $\Psi^{{\rm ii}} = (\delta W^+_\mu, \delta\phi_1)$ (i.e., $\Psi^{{\rm ii}}$ contains all Lorentz vector components of the $\delta W^+$ field and the upper scalar perturbations $\delta\phi_1$), and the operator acting on it is
\begin{equation}\label{eq:operDiiM}
\mathcal{D}^{{\rm ii}} =
 \begin{pmatrix}
    \mathcal{D}^{{\rm ii},\mu}_{1,1,\nu} & \mathcal{D}^{{\rm ii}}_{1,2,\nu}\\
    \mathcal{D}^{{\rm ii},\mu}_{2,1}     & \mathcal{D}^{{\rm ii}}_{2,2}
 \end{pmatrix}\,,
\end{equation}
with the matrix elements
\begin{equation}\label{eq:operDii}
 \begin{aligned}
  \mathcal{D}^{{\rm ii},\mu}_{1,1,\nu} &= \left[\square + g^2 \left(\frac{1}{2}\phi_2^*\phi_2-W_\rho^3 W^{\rho3}\right)-\imagi g\partial_\rho W^{\rho 3}-2\imagi g W^{\rho^3}\partial_\rho\right]\delta^\mu_\nu + 2\imagi g {\tilde W}^3_\nu{}^\mu\,,\\
  \mathcal{D}^{{\rm ii}}_{1,2,\nu} &= -2\imagi g D_\nu \phi_2\,,\\
  \mathcal{D}^{{\rm ii},\mu}_{2,1} &= -\sqrt{2}\imagi g D^\mu\phi_2\,,\\
  \mathcal{D}^{{\rm ii}}_{2,2} &= (\partial_\rho -\imagi g_{\scriptscriptstyle Z\phi^+} Z_\rho -\imagi g_{\scriptscriptstyle X\phi^+}X_\rho)^2 +  \beta_1 (|\phi_2|^2 -1 ) +\beta'(|\chi|^2 -\eta_2^2) +\frac{g^2}{2}|\phi_2|^2\,.\\
 \end{aligned}
\end{equation}
This is the block that is known to yield the negative eigenvalues corresponding to instabilities in the case of the electroweak string. Block (iii) is merely the complex conjugate of block (ii).

In block (iv), $\Psi^{{\rm iv}} = (\delta Z_\mu, \delta X_\mu, \delta\phi_2, \delta\phi_2^*, \delta\chi, \delta \chi^*)$, and the operator acting on it is
\begin{equation}\label{eq:operDivM}
 \mathcal{D}^{{\rm iv}} =
 \begin{pmatrix}
% \delta Z_\mu row
 \mathcal{D}^{{\rm iv},\mu}_{1,1,\nu} & \mathcal{D}^{{\rm iv},\mu}_{1,2,\nu} & \mathcal{D}^{{\rm iv}}_{1,3,\nu} & \mathcal{D}^{{\rm iv}}_{1,4,\nu} & \mathcal{D}^{{\rm iv}}_{1,5,\nu} & \mathcal{D}^{{\rm iv}}_{1,6,\nu} \\
%  \delta X_\mu row
 \mathcal{D}^{{\rm iv},\mu}_{2,1,\nu} & \mathcal{D}^{{\rm iv},\mu}_{2,2,\nu} & \mathcal{D}^{{\rm iv}}_{2,3,\nu} & \mathcal{D}^{{\rm iv}}_{2,4,\nu} & \mathcal{D}^{{\rm iv}}_{2,5,\nu} & \mathcal{D}^{{\rm iv}}_{2,6,\nu} \\
%  \delta \phi_2 row
 \mathcal{D}^{{\rm iv},\mu}_{3,1}  & \mathcal{D}^{{\rm iv},\mu}_{3,2} & \mathcal{D}^{{\rm iv}}_{3,3} & \mathcal{D}^{{\rm iv}}_{3,4} & \mathcal{D}^{{\rm iv}}_{3,5} & \mathcal{D}^{{\rm iv}}_{3,6}\\
%  \delta \phi_2^* row
 \mathcal{D}^{{\rm iv},\mu}_{4,1}  & \mathcal{D}^{{\rm iv},\mu}_{4,2} & \mathcal{D}^{{\rm iv}}_{4,3} & \mathcal{D}^{{\rm iv}}_{4,4} & \mathcal{D}^{{\rm iv}}_{4,5} & \mathcal{D}^{{\rm iv}}_{4,6}\\
%  \delta \chi row
 \mathcal{D}^{{\rm iv},\mu}_{5,1}  & \mathcal{D}^{{\rm iv},\mu}_{5,2} & \mathcal{D}^{{\rm iv}}_{5,3} & \mathcal{D}^{{\rm iv}}_{5,4} & \mathcal{D}^{{\rm iv}}_{5,5} & \mathcal{D}^{{\rm iv}}_{5,6}\\
%  \delta \chi^* row
 \mathcal{D}^{{\rm iv},\mu}_{6,1}  & \mathcal{D}^{{\rm iv},\mu}_{6,2} & \mathcal{D}^{{\rm iv}}_{6,3} & \mathcal{D}^{{\rm iv}}_{6,4} & \mathcal{D}^{{\rm iv}}_{6,5} & \mathcal{D}^{{\rm iv}}_{6,6}
 \end{pmatrix}\,,
\end{equation}
with the matrix elements
\begin{equation}\label{eq:operDiv}
 \begin{aligned}
% Z
\mathcal{D}^{{\rm iv},\mu}_{1,1,\nu} &= \left[\square + 2(g_{\scriptscriptstyle ZH}^2 \phi_2^*\phi_2 +g_{\scriptscriptstyle ZS}^2\chi^*\chi)\right]\delta^\mu_\nu\,,\\
% ZX, XZ
\mathcal{D}^{{\rm iv},\mu}_{1,2,\nu} &= D^{{\rm iv},\mu}_{2,1,\nu} = 2(g_{\scriptscriptstyle XH}|\phi_2|^2+g_{\scriptscriptstyle ZS}g_{\scriptscriptstyle XS}|\chi|^2)\delta^\mu_\nu\,,\\
% Zphi2, Zphi2*
\mathcal{D}^{{\rm iv}}_{1,3,\nu} &= \mathcal{D}^{{\rm iv}*}_{1,4,\nu} = -2\imagi g_{\scriptscriptstyle ZH} (D_\nu\phi_2)^*\,,\\
% Zchi, Zchi*
\mathcal{D}^{{\rm iv}}_{1,5,\nu} &= \mathcal{D}^{{\rm iv}*}_{1,6,\nu} = -2\imagi g_{\scriptscriptstyle ZS}(\tilde{D}_\nu \chi)^*\,,\\
% X
\mathcal{D}^{{\rm iv},\mu}_{2,2,\nu} &= \left[\square + 2(g_{\scriptscriptstyle XH}^2 \phi_2^*\phi_2 +g_{\scriptscriptstyle XS}^2\chi^*\chi)\right]\delta^\mu_\nu\,,\\
% Xphi2, Xphi2*
\mathcal{D}^{{\rm iv}}_{2,3,\nu} &= \mathcal{D}^{{\rm iv}*}_{2,4,\nu} = -2\imagi g_{\scriptscriptstyle XH} (D_\nu\phi_2)^*\,,\\
% Xchi, Xchi*
\mathcal{D}^{{\rm iv}}_{2,5,\nu} &= \mathcal{D}^{{\rm iv}*}_{2,6,\nu} = -2\imagi g_{\scriptscriptstyle XS}(\tilde{D}_\nu \chi)^*\,,\\
 \end{aligned}\quad\quad
\begin{aligned}
 % phi2 Z
 \mathcal{D}^{{\rm iv},\mu}_{3,1} &= \mathcal{D}^{{\rm iv},\mu*}_{4,1} = 2\imagi g_{\scriptscriptstyle ZH}D^\mu \phi_2\,,\\
 % phi2 X
 \mathcal{D}^{{\rm iv},\mu}_{3,2} &= \mathcal{D}^{{\rm iv},\mu*}_{4,2} =
 2\imagi g_{\scriptscriptstyle XH}D^\mu \phi_2\,,\\
 % phi2 phi2^*
 \mathcal{D}^{{\rm iv}}_{3,4}\, &= \mathcal{D}^{{\rm iv}*}_{4,3} = (\beta_1 - g_{\scriptscriptstyle ZH}^2 -g_{\scriptscriptstyle XH}^2)\phi_2^2\,,\\
 % phi2*
 \mathcal{D}^{{\rm iv}}_{4,4} &= \mathcal{D}^{{\rm iv}*}_{3,3}\,,\\
 % chi Z
 \mathcal{D}^{{\rm iv},\mu}_{5,1} &= \mathcal{D}^{{\rm iv},\mu*}_{6,1} = -2\imagi g_{\scriptscriptstyle ZS}{\tilde{D}}^\mu \chi\,,\\
 %  chi X
 \mathcal{D}^{{\rm iv},\mu}_{5,2} &= \mathcal{D}^{{\rm iv},\mu*}_{6,2} = -2\imagi g_{\scriptscriptstyle XS}{\tilde{D}}^\mu \chi\,,\\
 % chi*
 \mathcal{D}^{{\rm iv}}_{6,6}\, &= \mathcal{D}^{{\rm iv}*}_{5,5}\,,\\
\end{aligned}
\end{equation}
and
\begin{equation}\label{eq:operDiv2}
 \begin{aligned}
 % phi2
 \mathcal{D}^{{\rm iv}}_{3,3} &= (\partial_\mu -\imagi g_{\scriptscriptstyle ZH} Z_\mu -\imagi g_{\scriptscriptstyle XH}X_\mu)^ 2+ \beta_1 (2|\phi_2|^2 - 1 ) +\beta'(|\chi|^2 -\eta_2^2) + (g_{\scriptscriptstyle ZH}^2+g_{\scriptscriptstyle XH}^2)|\phi_2|^2\,,\\
 % phi2 chi, phi2* chi*, chi phi2, chi* phi2*
 \mathcal{D}^{{\rm iv}}_{3,5} &= \mathcal{D}^{{\rm iv}}_{5,3} = \mathcal{D}^{{\rm iv}*}_{4,6} = \mathcal{D}^{{\rm iv}*}_{6,4}= (\beta' + g_{\scriptscriptstyle ZH}g_{\scriptscriptstyle ZS} + g_{\scriptscriptstyle XH} g_{\scriptscriptstyle XS})\phi_2 \chi^*\,,\\
 % phi2 chi*, phi2* chi, chi* phi2, chi phi2*
 \mathcal{D}^{{\rm iv}}_{3,6} &= \mathcal{D}^{{\rm iv}}_{6,3} = \mathcal{D}^{{\rm iv}*}_{4,5} = \mathcal{D}^{{\rm iv}*}_{5,4}= (\beta' - g_{\scriptscriptstyle ZH}g_{\scriptscriptstyle ZS} - g_{\scriptscriptstyle XH} g_{\scriptscriptstyle XS})\phi_2 \chi^*\,,\\
 % chi
 \mathcal{D}^{{\rm iv}}_{5,5} &= (\partial_\rho-\imagi g_{\scriptscriptstyle ZS}Z_\rho -\imagi g_{\scriptscriptstyle XS}X_\rho)^2 + \beta_2 (2|\chi|^2-\eta_2^2 ) +\beta'(|\phi_2|^2-1)+(g_{\scriptscriptstyle ZS}^2+g_{\scriptscriptstyle XS}^2)|\chi|^2\,.\\
 \end{aligned}
\end{equation}
Note that in Eq.\ (\ref{eq:operDiv2}), in the first bracketed term of $\mathcal{D}^{{\rm iv}}_{4,4}$ the square represents a contraction over the index $\rho$.

In order to bring Eq.\ (\ref{eq:lineq}) to a form tractable numerically, we shall consider the Fourier transform in the coordinated $z$ and $t$, and note that Fourier components are decoupled, apart from ones corresponding to $k$ and $-k$, $\Omega$, and $-\Omega$,
\begin{equation}
 \label{eq:Fourier}
 \Psi (x_i,z,t) = \Phi(x,\Omega, k) \exp[\imagi(\Omega t-k z)]\,,
\end{equation}
where $i=1,2$, and the components of the Fourier transformed field are $\Phi(x,\Omega, k)=(\delta \tilde A_\mu$, $\delta \tilde W^+_\mu$,$\delta \tilde W^-_\mu$, $\delta \tilde Z_\mu$, $\delta \tilde X_\mu$, $\delta \tilde \phi_a$, $\delta\tilde \phi_a^*$, $\delta\tilde \chi$, $\delta\tilde \chi^*)$, depending on the variables $(r, \vartheta, \Omega, k)$.

We also apply a partial wave decomposition in the angular coordinate $\vartheta$ to the components of $\Phi(x,\Omega,k)$,
\begin{equation}
 \label{eq:Pwaves}
 \begin{aligned}
  \delta\tilde \phi_1 &= \e^{\imagi\ell\vartheta}s_{1,\ell}(r)\,,\\
  \delta\tilde \phi_2 &= \e^{\imagi(n+\ell)\vartheta}s_{2,\ell}(r)\,,\\
  \delta\tilde \chi   &= \e^{\imagi\ell\vartheta}s_{3,\ell}(r)\,,\\
  \delta\tilde Z_+   &= \e^{\imagi(\ell-1)\vartheta} \imagi z_\ell(r)\,,\\
  \delta\tilde X_+   &= \e^{\imagi(\ell-1)\vartheta} \imagi x_\ell(r)\,,\\
  \delta\tilde W^\pm_+ &= \e^{\imagi (\ell-1\mp n)\vartheta} \imagi w_{\pm,\ell}(r)\,,\\
  \delta\tilde Z_{3,4}&= \e^{\imagi \ell \vartheta}z_{3,4,\ell}(r)\,,\\
  \delta\tilde W^+_{3,4} &= \e^{\imagi \ell \vartheta}w_{3,4,\ell}(r)\,,
 \end{aligned}\quad\quad
 \begin{aligned}
  \delta\tilde \phi_1^* &= \e^{\imagi\ell\vartheta}s_{1,-\ell}^*(r)\,,\\
  \delta\tilde \phi_2^* &= \e^{-\imagi(n-\ell)\vartheta}s_{2,-\ell}^*(r)\,,\\
  \delta\tilde \chi^*   &= \e^{\imagi\ell\vartheta}s_{3,-\ell}^*(r)\,,\\
  \delta\tilde Z_-     &= \e^{\imagi(\ell+1)\vartheta} (-\imagi) z_{-\ell}^*(r)\,,\\
  \delta\tilde X_-     &= \e^{\imagi(\ell+1)\vartheta} (-\imagi) x_{-\ell}^*(r)\,,\\
  \delta\tilde W^\pm_- &= \e^{\imagi (\ell+1\mp n)\vartheta} (-\imagi) w_{\pm,-\ell}^*(r)\,,\\
  \delta\tilde X_{3,4} &= \e^{\imagi \ell \vartheta}x_{3,4,\ell}(r)\,,\\
  \delta\tilde W^-_{3,4} &= \e^{\imagi \ell \vartheta}w^*_{3,4,-\ell}(r)\,,
 \end{aligned}
\end{equation}
where $\delta\tilde Z_+ = \exp(-\imagi\vartheta)(\delta\tilde Z_r-\imagi\delta\tilde Z_\vartheta/r)$, and analogously for the other gauge fields. On the radial functions, the variables $\Omega$ and $k$ have been suppressed. In all equations these appear as $\Omega^2-k^2$, and therefore, the lowest eigenvalue corresponds to $k=0$; for this reason, $k$ is dropped in what follows.

In addition to the block structure of the time-dependent linearised equations (\ref{eq:lineq}), there is a further decoupling due to the time- and $z$-independence of the background solution (\ref{eq:Ansatz1}), (\ref{eq:Ansatz2}), resulting in a further decoupling of the $z$ and $t$ (0 and 3) components of the vector fields. The following blocks decouple and can be solved separately: (i) $\delta A_i$ ($i=1,2$); (ii) $\delta W^\pm$, $\delta\phi_1$; (iii) $\delta W^{\pm *}$, $\delta\phi_1^*$ [conjugate of (iii)]; (iv) $\delta Z_i$, $\delta X_i$, $\delta\phi_2$, $\delta\phi_2^*$ $\delta\chi$, $\delta\chi^*$; (v) $\delta A_3$; (vi) $\delta A_0$; (vii) $\delta Z_3$, $\delta X_3$; (viii) $\delta Z_0$, $\delta X_0$; (ix) $W^{\pm}_3$; and (x) $W^{\pm}_0$. %We note, that for $n=1$ electroweak vortices, the instabilities are in the sector (ii), therefore we shall pursue that block further.
%The explanation of this block structure is that for the background solution, a gauge was chosen in which it has a global direction in internal space, such that $\phi_1=0$, and the solution is time and $z$-independent. Also, the equations for the $z$ and the time component of the vector field perturbations agree.

Eigenvectors and eigenvalues in each block can be considered separately; therefore, we shall write the radial equations in block $I=i, \dots, x$ separately, in the form
\begin{equation}\label{eq:radeqEig}
 \mathcal{M}_\ell^I \Phi_\ell^I = \Omega^2 \Phi_\ell^I\,,
\end{equation}
with the block containing the known  instabilities of electroweak strings consisting of $\Phi_\ell^{\rm ii}=(s_{1,\ell}, w^+_{+,\ell}$, $w^+_{-,\ell})$. In this sector, the radial equations (\ref{eq:radeqPMM}) are obtained, with the index $I={\rm ii}$ dropped, and this block is considered in detail in Sec.\ \ref{sec:stab}, where its numerical solution is also discussed. Block (iii) contains the same equations for the complex conjugates $s_{1,-\ell}^*$, $w^{-*}_{-,-\ell}$, $w^{-*}_{+,-\ell}$, with the replacement $\ell\to -\ell$, as block (ii).

Of the remaining blocks,  (i), (v), and (vi) merely contain the radial Laplacian. Block (iv) contains a deformation of the eigenvalue problem of the ANO string (or equivalently, that of the semilocal-dark string \cite{FLS}), and possesses only positive eigenvalues: $\Phi^{\rm iv}_\ell = (s_{2,\ell}$, $s_{2,-\ell}^*$, $s_{3,\ell}$, $s_{3,-\ell}^*$, $z_\ell$, $z_{-\ell}^*$, $x_\ell$, $x_{-\ell}^*)$, and the elements of the corresponding operator $\mathcal{M}_\ell^{\rm iv}$ are
\begin{equation}\label{eq:operMliv}
 \begin{aligned}
  \mathcal{M}_{\ell,1,1}^{\rm iv} &= -\frac{\d^2}{\d r^2}-\frac{1}{r}\frac{\d}{\d r}\\ &\,+\left[ \frac{[\ell+n(1-\mathfrak{z}-g_{\scriptscriptstyle XH}x)]^2}{r^2}+\beta_1(2 f^2-1)+\beta'(f_d^2-\eta_2^2)+(1+g_{\scriptscriptstyle XH}^2)f^2\right]\,,\\
%  {\mathcal{M}_\ell^{\rm iv}}_{2,2} &= -\frac{\d^2}{\d r^2}-\frac{1}{r}\frac{\d}{\d r}+\left[ \frac{[\ell-n(1-z-g_{\scriptscriptstyle XH}x)]^2}{r^2}+\beta_1(2 f^2-1)+\beta'(f_d^2-\eta_2^2)+(1+g_{\scriptscriptstyle XH}^2)f^2\right]\,,\\
  \mathcal{M}_{\ell,3,3}^{\rm iv} &= -\frac{\d^2}{\d r^2}-\frac{1}{r}\frac{\d}{\d r}\\&\,+\left[ \frac{[\ell-n(g_{\scriptscriptstyle ZS}\mathfrak{z}+g_{\scriptscriptstyle XS}x)]^2}{r^2}+\beta_2(2 f_d^2-\eta_2^2)+\beta'(f^2-1)+(g_{\scriptscriptstyle ZS}^2+g_{\scriptscriptstyle XS}^2)f_d^2\right]\,,\\
  \mathcal{M}_{\ell,5,5}^{\rm iv} &= -\frac{\d^2}{\d r^2}-\frac{1}{r}\frac{\d}{\d r}+\left[ \frac{(\ell-1)^2}{r^2}+2(f^2 + g_{\scriptscriptstyle ZS}^2 f_d^2)\right]\,,\\
  \mathcal{M}_{\ell,7,7}^{\rm iv} &= -\frac{\d^2}{\d r^2}-\frac{1}{r}\frac{\d}{\d r}+\left[ \frac{(\ell-1)^2}{r^2}+2(g_{\scriptscriptstyle XH}^2f^2 + g_{\scriptscriptstyle XS}^2 f_d^2)\right]\,,\\
\end{aligned}
\end{equation}
and
\begin{equation}\label{eq:additMliv}
 \begin{aligned}
  \mathcal{M}_{\ell,2,2}^{\rm iv} &= \mathcal{M}_{-\ell,1,1}^{\rm iv}\,,\\
  \mathcal{M}_{\ell,6,6}^{\rm iv} &= \mathcal{M}_{-\ell,5,5}^{\rm iv}\,,
 \end{aligned}\quad\quad
 \begin{aligned}
  \mathcal{M}_{\ell,4,4}^{\rm iv} &= \mathcal{M}_{-\ell,3,3}^{\rm iv}\,,\\
  \mathcal{M}_{\ell,8,8}^{\rm iv} &= \mathcal{M}_{-\ell,7,7}^{\rm iv}\,,
 \end{aligned}
\end{equation}
with the couplings
\begin{equation}\label{eq:couplMliv}
 \begin{aligned}
 % s2l, s2*-l
  \mathcal{M}_{\ell,1,2}^{\rm iv} &= \mathcal{M}_{\ell,2,1}^{\rm iv} = (\beta_1-1-g_{\scriptscriptstyle XH}^2)f^2\,,\\
 % s2l, s3l
  \mathcal{M}_{\ell,1,3}^{\rm iv} &= \mathcal{M}_{\ell,3,1}^{\rm iv} = \mathcal{M}_{\ell,2,4}^{\rm iv} = \mathcal{M}_{\ell,4,2}^{\rm iv} = (\beta'+g_{\scriptscriptstyle ZS}+g_{\scriptscriptstyle XH}g_{\scriptscriptstyle XS})ff_d\,,\\
 % s2l, s3*-l
  \mathcal{M}_{\ell,1,4}^{\rm iv} &= \mathcal{M}_{\ell,4,1}^{\rm iv} = \mathcal{M}_{\ell,2,3}^{\rm iv} = \mathcal{M}_{\ell,3,2}^{\rm iv} = (\beta'-g_{\scriptscriptstyle ZS}-g_{\scriptscriptstyle XH}g_{\scriptscriptstyle XS})ff_d\,,\\
 % s2l, zl
  \mathcal{M}_{\ell,1,5}^{\rm iv} &= \mathcal{M}_{\ell,5,1}^{\rm iv} = \mathcal{M}_{\ell,2,6}^{\rm iv} = \mathcal{M}_{\ell,6,2}^{\rm iv} = - \sqrt{2}\left(f'-\frac{nf}{r}(1-\mathfrak{z}-g_{\scriptscriptstyle XH}x)\right)\,,\\
 % s2l, z*-l
  \mathcal{M}_{\ell,1,6}^{\rm iv} &= \mathcal{M}_{\ell,6,1}^{\rm iv} = \mathcal{M}_{\ell,2,5}^{\rm iv} = \mathcal{M}_{\ell,5,2}^{\rm iv} = \sqrt{2}\left(f'+\frac{nf}{r}(1-\mathfrak{z}-g_{\scriptscriptstyle XH}x)\right)\,,\\
 % s2l, xl
  \mathcal{M}_{\ell,1,7}^{\rm iv} &= \mathcal{M}_{\ell,7,1}^{\rm iv} = \mathcal{M}_{\ell,2,8}^{\rm iv} = \mathcal{M}_{\ell,8,2}^{\rm iv} = g_{\scriptscriptstyle XS} \mathcal{M}_{\ell,1,5}^{\rm iv}\,,\\
 % s2l, x*-l
  \mathcal{M}_{\ell,1,8}^{\rm iv} &= \mathcal{M}_{\ell,8,1}^{\rm iv} = \mathcal{M}_{\ell,2,7}^{\rm iv} = \mathcal{M}_{\ell,7,1}^{\rm iv} = g_{\scriptscriptstyle XS} \mathcal{M}_{\ell,1,6}^{\rm iv}\,,\\
 % s3l s3*-l
  \mathcal{M}_{\ell,3,4}^{\rm iv} &= \mathcal{M}_{\ell,4,3}^{\rm iv} = (\beta_2 -g_{\scriptscriptstyle ZS}^2-g_{\scriptscriptstyle XS}^2)f_d^2\,,\\
 % s2l zl
  \mathcal{M}_{\ell,3,5}^{\rm iv} &= \mathcal{M}_{\ell,5,3}^{\rm iv} = \mathcal{M}_{\ell,4,6}^{\rm iv} = \mathcal{M}_{\ell,6,4}^{\rm iv} = -\sqrt{2} \left( f_d' +\frac{nf_d}{r}(g_{\scriptscriptstyle XS}\mathfrak{z}+g_{\scriptscriptstyle XS}x)\right)g_{\scriptscriptstyle ZS}\,,\\
 % s2l z*-l
  \mathcal{M}_{\ell,3,6}^{\rm iv} &= \mathcal{M}_{\ell,6,3}^{\rm iv} = \mathcal{M}_{\ell,4,5}^{\rm iv} = \mathcal{M}_{\ell,5,4}^{\rm iv} = \sqrt{2} \left( f_d' -\frac{nf_d}{r}(g_{\scriptscriptstyle XS}\mathfrak{z}+g_{\scriptscriptstyle XS}x)\right)g_{\scriptscriptstyle ZS}\,,\\
 % s2l xl
  \mathcal{M}_{\ell,3,7}^{\rm iv} &= \mathcal{M}_{\ell,7,3}^{\rm iv} = \mathcal{M}_{\ell,4,8}^{\rm iv} = \mathcal{M}_{\ell,8,4}^{\rm iv} = -\sqrt{2} \left( f_d' +\frac{nf_d}{r}(g_{\scriptscriptstyle XS}\mathfrak{z}+g_{\scriptscriptstyle XS}x)\right)g_{\scriptscriptstyle XS}\,,\\
 % s2l x*-l
  \mathcal{M}_{\ell,3,8}^{\rm iv} &= \mathcal{M}_{\ell,8,3}^{\rm iv} = \mathcal{M}_{\ell,4,7}^{\rm iv} = \mathcal{M}_{\ell,7,4}^{\rm iv} = \sqrt{2} \left( f_d' -\frac{nf_d}{r}(g_{\scriptscriptstyle XS}\mathfrak{z}+g_{\scriptscriptstyle XS}x)\right)g_{\scriptscriptstyle XS}\,,\\
 % z x
  \mathcal{M}_{\ell,5,7}^{\rm iv} &= \mathcal{M}_{\ell,7,5}^{\rm iv} = \mathcal{M}_{\ell,6,8}^{\rm iv} = \mathcal{M}_{\ell,8,6}^{\rm iv} = 2(g_{\scriptscriptstyle XH}f^2 + g_{\scriptscriptstyle ZS}g_{\scriptscriptstyle XS} f_d^2)\,.
 \end{aligned}
\end{equation}

The equations in blocks (vii) and (viii) are identical. Let now $\Phi_\ell^{\rm vii} = (z_{3\ell}, x_{3\ell})$, and
\begin{equation}\label{eq:operMlvii}
 \begin{aligned}
  \mathcal{M}_{\ell,1,1}^{\rm vii} &=  -\frac{\d^2}{\d r^2}-\frac{1}{r}\frac{\d}{\d r}+\left[ \frac{\ell^2}{r^2}+2f^2 + 2g_{\scriptscriptstyle ZS}^2 f_d^2)\right]\,,\\
  \mathcal{M}_{\ell,2,2}^{\rm vii} &=  -\frac{\d^2}{\d r^2}-\frac{1}{r}\frac{\d}{\d r}+\left[ \frac{\ell^2}{r^2}+2g_{\scriptscriptstyle XH}^2f^2 + 2g_{\scriptscriptstyle XS}^2 f_d^2)\right]\,,\\
  \mathcal{M}_{\ell,1,2}^{\rm vii} &= \mathcal{M}_{\ell,2,1}^{\rm vii} = 2(g_{\scriptscriptstyle XH}f^2 + g_{\scriptscriptstyle ZS}g_{\scriptscriptstyle XS}f_d^2)\,.
 \end{aligned}
\end{equation}

In block (ix), $w_{3,\ell}$ and $w_{3,-\ell}^*$ decouple, the equation for the former is
\begin{equation}\label{eq:deltawz}
 -\frac{1}{r}(rw_{3,\ell}')' +\left[ \frac{[\ell-g n(\alpha_2 \mathfrak{z}+\alpha_3 x)^2]}{r^2}+\frac{g}{2}f^2\right]w_{3,\ell} = \Omega^2 w_{3,\ell}\,,
\end{equation}
and the equation for $w_{3,-\ell}^*$ is obtained by the replacement $\ell \to -\ell$, $w_{3,\ell}\to w_{3,-\ell}$.

The remaining gauge freedom is characterised by ghost equations: an infinitesimal gauge transformation substituted into Eq.\ (\ref{eq:lineq}). The general form of an infinitesimal gauge transformation is
\begin{equation}\label{eq:infGT}
 \begin{aligned}
  \delta\phi_1 &= \imagi(g_{\scriptscriptstyle Z\phi^+} \xi_{\scriptscriptstyle Z}+g_{\scriptscriptstyle X\phi^+}\xi_{\scriptscriptstyle X})\phi_1 +\frac{\imagi g}{2}\xi^+ \phi_2\,,\\
  \delta\phi_2 &= \imagi(\xi_{\scriptscriptstyle Z}+g_{\scriptscriptstyle XH}\xi_{\scriptscriptstyle X})\phi_2 +\frac{\imagi g}{2}\xi^- \phi_1\,,\\
  \delta\chi &= \imagi(g_{\scriptscriptstyle ZS} \xi_{\scriptscriptstyle Z}+g_{\scriptscriptstyle XS}\xi_{\scriptscriptstyle X})\chi\,,\\
  \delta Z_\mu &= \partial_\mu \xi_{\scriptscriptstyle Z}\,,\\
  \delta X_\mu &= \partial_\mu \xi_{\scriptscriptstyle X}\,,\\
  \delta W^+_\mu &= \partial_\mu \xi^+ -\imagi g W^3_\mu \xi^+\,,\\
  \delta W^-_\mu &= \partial_\mu \xi^- +\imagi g W^3_\mu \xi^-\,,
 \end{aligned}
\end{equation}
where the functions $\xi_{\scriptscriptstyle Z}$, $\xi_{\scriptscriptstyle X}$, and $\xi^\pm$ are generators of the infinitesimal gauge transformations. The radial ghost equations for the Fourier coefficients of these functions are as follows:
%The ghost equation in sector $ii$ is
\begin{equation}
 \label{eq:ghost}
 \begin{aligned}
 -\frac{1}{r}(r\xi_{{\scriptscriptstyle Z}\ell}')' +\left[ \frac{\ell^2}{r^2}+2(f^2+g_{\scriptscriptstyle ZS}^2f_d^2)\right] \xi_{{\scriptscriptstyle Z}\ell}+2(g_{\scriptscriptstyle XH}f^2+g_{\scriptscriptstyle ZS}g_{\scriptscriptstyle XS}f_d^2)\xi_{{\scriptscriptstyle X}\ell} &= \Omega^2 \xi_{{\scriptscriptstyle Z}\ell}\,,\\
 -\frac{1}{r}(r\xi_{{\scriptscriptstyle X}\ell}')' +\left[ \frac{\ell^2}{r^2}+2(g_{\scriptscriptstyle XH}^2f^2+g_{\scriptscriptstyle XS}^2f_d^2)\right] \xi_{{\scriptscriptstyle X}\ell}+2(g_{\scriptscriptstyle XH}f^2+g_{\scriptscriptstyle ZS}g_{\scriptscriptstyle XS}f_d^2)\xi_{{\scriptscriptstyle Z}\ell} &= \Omega^2 \xi_{X\ell}\,,\\
 -\frac{1}{r}(r{\xi^+_\ell}')'+\left[ \frac{[\ell-ng(\alpha_2 \mathfrak{z}+\alpha_3 x)]^2}{r^2}+\frac{g}{2}f^2\right]\xi^+_\ell&=\Omega^2 \xi^+_\ell\,,\\
 -\frac{1}{r}(r{\xi^-_\ell}')'+\left[ \frac{[\ell+ng(\alpha_2 \mathfrak{z}+\alpha_3 x)]^2}{r^2}+\frac{g}{2}f^2\right]\xi^-_\ell&=\Omega^2 \xi^-_\ell\,,
 \end{aligned}
\end{equation}
which are all deformations of the ghost equation for ANO, semilocal, or semilocal-dark vortex ghost equations, which all have relatively large positive eigenvalues \cite{Goodband, FLCC, FLCC2, FLS}, therefore, they are not required for stability analysis.

\def\refttl#1{{\sl ``#1''}, }%

\end{document}